\documentclass[sigplan,nonacm]{acmart}
\usepackage{graphicx} 
\usepackage{tikzit}

\tikzstyle{Z dot}=[draw=black, shape=circle, fill={rgb,255: red,216; green,248; blue,216}, inner sep=0.7mm, minimum width=0pt, minimum height=0pt]
\tikzstyle{Z phase dot}=[shape=rectangle, fill={rgb,255: red,216; green,248; blue,216}, draw=black, minimum size=1em, rounded corners=0.4em, inner sep=0.2em, outer sep=-0.2em, scale=0.8, font={\scriptsize}]
\tikzstyle{X dot}=[Z dot, fill={rgb,255: red,232; green,165; blue,165}, draw=black]
\tikzstyle{X phase dot}=[Z phase dot, fill={rgb,255: red,232; green,165; blue,165}, draw=black]
\tikzstyle{hadamard}=[draw=black, shape=rectangle, fill=yellow, inner sep=0.6mm, minimum height=1.5mm, minimum width=1.5mm]
\tikzstyle{Z box}=[draw=black, shape=rectangle, fill={rgb,255: red,216; green,248; blue,216}, minimum size=.55em, inner sep=0.15em, scale=0.85, font={\scriptsize}]
\tikzstyle{W triangle}=[shape=isosceles triangle, isosceles triangle stretches=true, fill=black, draw=black, minimum width=2.8mm, minimum height=2mm, inner sep=1pt, outer sep=0pt, shape border rotate=90]
\tikzstyle{W input}=[fill=none, inner sep=0mm, outer sep=0mm, minimum width=0mm]
\tikzstyle{label}=[inner sep=0mm, font={\small}]

\tikzstyle{hadamard edge}=[-, dashed, dash pattern=on 2pt off 1.5pt, thick, draw={rgb,255: red,68; green,136; blue,255}]
\tikzstyle{W io edge}=[-, draw=black]

\usepackage{siunitx}
\usepackage{outlines}
\usepackage{multirow}
\usepackage{svg}
\usepackage{braket}
\usepackage{makecell}
\usepackage{booktabs}
\usepackage{mathtools}
\DeclarePairedDelimiter{\ceil}{\lceil}{\rceil}
\title{Architecting Distributed Quantum Computers: 
\\
Design Insights from Resource Estimation}

\author{Dmitry Filippov}
\affiliation{%
  \institution{Department of Computer Science and Technology, University of Cambridge}
  \country{United Kingdom}
}

\author{Peter Yang}
\affiliation{%
  \institution{Cavendish, Department of Physics, University of Cambridge}
  \country{United Kingdom}
}

\author{Prakash Murali}
\affiliation{%
  \institution{Department of Computer Science and Technology, University of Cambridge}
  \country{United Kingdom}
}
\date{May 2025}

\begin{document}

\begin{abstract}

In the emerging field of Fault Tolerant Quantum Computation (FTQC), resource estimation is an important tool for quantitatively comparing prospective architectures, identifying hardware bottlenecks and informing which research paths are most valuable. Despite a recent increase in attention on FTQC, there is currently a lack of resource estimation research for architectures that can realistically offer quantum advantage. In particular, current modelling efforts focus on \emph{monolithic} quantum computers where all qubits reside on a single device. Constraints on fabrication yield, wiring density, and cooling power make monolithic devices unlikely to scale to fault-tolerant sizes in the foreseeable future. \emph{Distributed quantum supercomputers} offer a path to overcome these limitations. 

We propose a prospective distributed quantum computing architecture based on lattice surgery with support for modular and distributed operations, with a focus on superconducting qubits. 
We develop a resource-estimation framework and software tool tailored to distributed FTQC, enabling end-to-end analysis of practical quantum algorithms on our proposed architecture with various hardware configurations, spanning different node sizes, inter-node entanglement generation rates and distillation protocols. Our extensive benchmarking across eight applications and thousands of hardware configurations, shows that resource estimation driven architecture design is crucial for scalability. We provide concrete design configurations that have feasible resource requirements, recommendations for hardware design and system organization. More broadly, our work provides a rigorous methodology for architectural pathfinding, capable of informing system designs and guiding future research priorities.

The tooling used in this paper will be made available on publication with an open source and will enable future benchmarking and resource estimation studies.

\end{abstract}

\maketitle
\section{Introduction}


Quantum computing is expected to aid in solving problems intractable by classical means, such as factoring \cite{shor_factoring, gidney2021factor} and quantum chemistry simulations \cite{quantum_catalysis, childs2018toward}.
 
Noisy quantum devices are now operational in the hundred to thousand qubit scale \cite{IBM, PhysRevX.13.041052, preskill2025beyond}. Experiments have also demonstrated quantum error correction (QEC), with systems that have few tens of \emph{logical qubits} that are composed of collections of physical qubits~\cite{Bluvstein2024, google2025quantum}. 
This marks the culmination of the Noisy Intermediate-Scale Quantum (NISQ) era and a gradual move towards Fault-Tolerant Quantum Computing (FTQC). To achieve quantum advantage over classical computing, FTQC systems need to scale up to $50$-$100$K physical qubits for scientifically interesting applications and to 1 million qubits for commercially relevant applications~\cite{Beverland2022-kd, gidney2021factor, 2411.10406}.

Current quantum systems are based on a \emph{monolithic} architecture where all qubits reside in a single physical device. This architecture faces severe scaling issues. First, for superconducting qubits yield (the percentage of functional qubits among manufactured qubits) is less than 50\%  and worsens with device size~\cite{VanDamme2024, Hertzberg2021}. Chiplet-based solutions alleviate yield issues \cite{kate_chiplet}, but still offer only 80\% yield. Second, superconducting qubits are large, occupying more than $1\mathrm{mm}^2$ device area per qubit\cite{Arute2019, 2501.04612}. A million-qubit device would need a wafer size of $1\mathrm{m}^2$, which is beyond all known manufacturing capabilities. Third, several wires are typically required per qubit from classical control electronics down to the quantum chip~\cite{Arute2019}. At the million-qubit scale, such wiring requirements are impossible to realize in practice~\cite{2411.10406}. Similar challenges exist in trapped ion and neutral atom systems where qubit loss during operation and the size of control electronics limits scaling~\cite{trapped_ion_blueprint, Bluvstein2024}. These challenges of yield, size and controllability mean that monolithic architectures cannot scale to sizes needed for quantum advantage.

These considerations suggest that the most viable route to large-scale quantum computation is through \emph{distributed quantum computing architectures}, in which multiple \emph{nodes} (small quantum processors) are networked together. It is similar to a classical supercomputer with compute nodes connected by a high-performance interconnect.  In the last few years, network components required for realizing distributed quantum computing have been pursued by the superconducting, trapped ion, silicon spin and neutral atom communities~\cite{photonic_paper, Sunami2025, Chou2018, 2302.08756}. This year, the first distributed trapped ion system with a photonic quantum interconnect was experimentally demonstrated~\cite{Main2025}. 

Industry and national roadmaps have also begun recognizing distributed computing as the path forward~\cite{nqcc, IBM, DARPA}.

Resource estimation and architecture modelling are key enablers of technological progress, allowing us to optimize future system designs and inform hardware roadmaps.  
Several resource estimation frameworks exist for monolithic quantum systems, including Qualtran \cite{harrigan2024qualtran} and the Azure Quantum Resource Estimator \cite{Beverland2022-kd}. 
These tools do not model the quantum network or mechanisms for \emph{entanglement distillation} (conversion of noisy Bell pairs generated by hardware into usable high-fidelity Bell pairs) that are necessary to build reliable components using noisy communication hardware. Our work shows these works significantly underestimate the resource needs of applications on distributed architectures. 
Other efforts~\cite{2411.10406, IBM} discuss distributed quantum architecture, but do not offer application-based architectural analysis across hardware configurations.

Our work aims to characterize the resource requirements of distributed quantum systems for practically useful, large-scale applications. We focus primarily on Lattice-Surgery based superconducting computers as their fast clock speeds allow them to run commercially interesting algorithms such as Shor's algorithm and molecular ground state estimation in a reasonable time frame\cite{Beverland2022-kd} of days. Neutral atom and trapped ion based quantum computers, in comparison, have operation times that are several orders of magnitude longer, making them less relevant to these applications. Despite not being optimised with these slower architectures in mind, our resource estimation framework can still model their operation assuming they are being used to implement lattice surgery based computation, as opposed to using other primitives such as transversal gates.

\begin{figure}[t]
    \centering
    \includegraphics[width=0.6\linewidth]{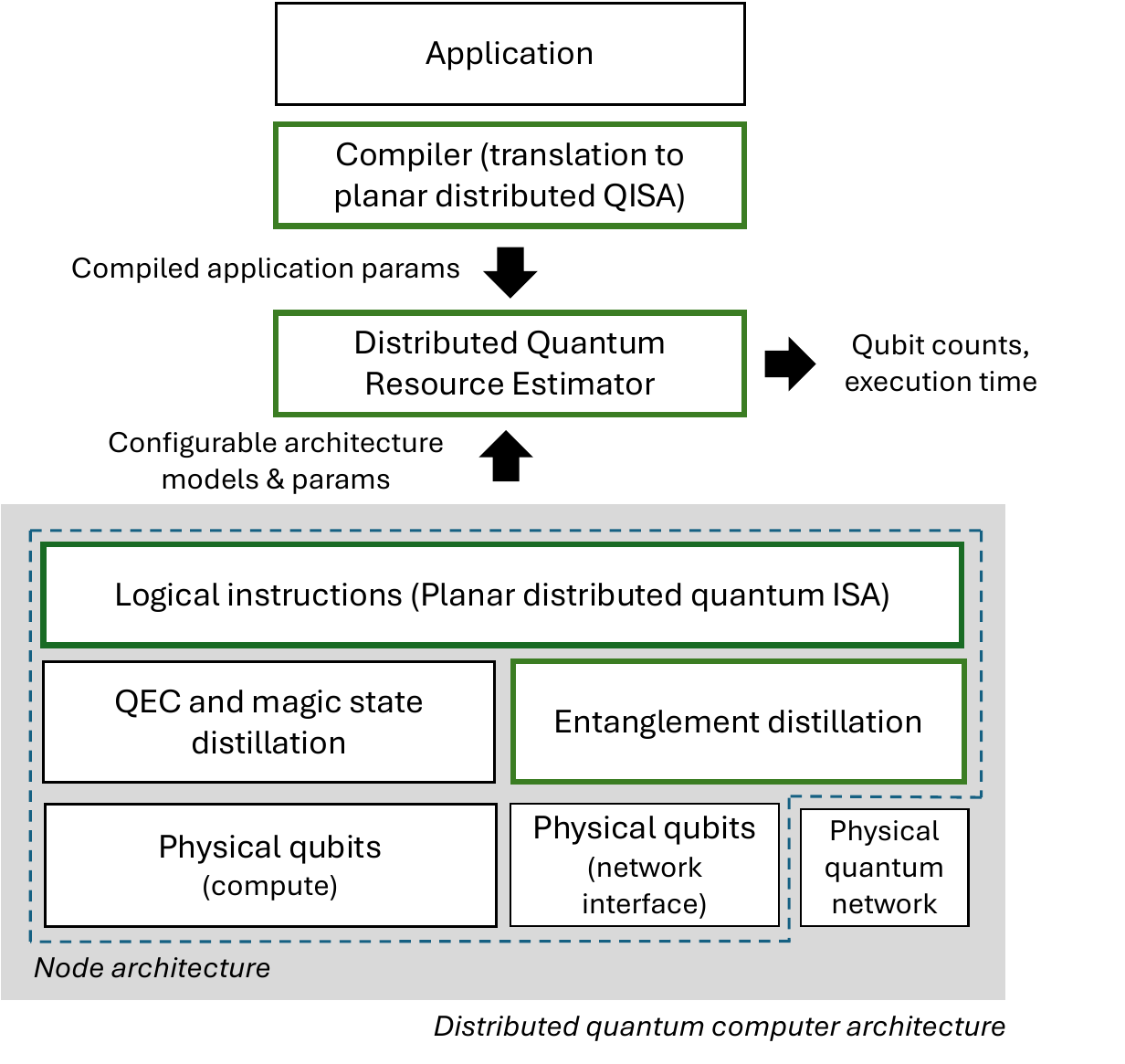}
    \caption{System architecture for the distributed quantum computer and our tool and modelling contributions highlighted in green. At the bottom, the grey box shows the overall system architecture consisting of nodes and the physical quantum network (mechanism for generating shared entanglement between nodes). The dashed box shows the architecture of a single node, with physical qubit, distillation and instruction set implementations. At the top, quantum applications are translated using a compiler to the instruction set.  Our core resource estimator is described in \S 7. It relies on the overall architecture and distributed planar quantum ISA developed in \S 3 and \S 4, models for the instruction translation in \S 5 and entanglement distillation modelling in \S 6.}
    \label{fig:architecture}
\end{figure}

We address three main challenges. First, parametrisable models that capture the qubit and time costs of entanglement distillation are lacking. Second, existing compilation techniques are tailored to monolithic devices and must be reformulated for distributed systems. Third, there is no unified resource-estimation framework that jointly optimizes resource allocation across the different components of the distributed system. To close these gaps, we developed a resource estimation framework (Figure \ref{fig:architecture}) designed specifically for distributed quantum architectures.  To our knowledge, it is the first to quantitatively integrate network-level processes with a full model of the fault-tolerant stack.
\\
\\
\\
\\
Our contributions are:

\begin{itemize} 
    \item A novel resource estimation framework designed for distributed quantum systems based on superconducting qubit platform. The framework takes  network and hardware parameters and algorithm resource counts as an input and produces the number of nodes and runtime required for space-time optimal execution of the algorithm on our proposed architecture. To do so, we co-optimise the algorithm's need for error correction, networking and magic state generation
    \item A framework for modelling performance of different error correction codes for the purposes of entanglement distillation (quantum network) on noisy hardware.
    \item A novel compilation technique that allows us to implement algorithms expressed in terms of Multi-Qubit Pauli Rotations (instruction set commonly used for lattice-surgery-based architectures \cite{Litinski2019, harrigan2024qualtran, Beverland2022-kd}) on a distributed quantum network using only local operations and logical bell state generation.
    
\end{itemize}

Using our resource estimation tool, we conduct a benchmarking study of practical quantum applications across a range of options for entanglement generation and distillation protocols. We answer design questions related to the hardware properties necessary for scaling, system organization and overall architecture prospects. Our design insights are:

\begin{itemize}
    \item \textbf{Feasibility:} Distributed architectures have practical resource requirements. Application resource needs are higher compared to monolithic architectures, but they are not prohibitively resource-intensive. 
    \item \textbf{Node and network architecture:} Based on our study and insights we propose an architecture capable of supporting the execution of arbitrary quantum algorithms given nodes of a fixed size (Figure \ref{fig:node-anatomy}).
    \item \textbf{Error rate:} Physical qubit error rates affect not only the size and runtime of the "local" part of the computation, but also the proportion of each node that needs to be dedicated to networking. Error rates of $10^{-4}$ or less are important for practically-feasible distributed quantum computation.
    \item \textbf{Network parameters:}  
    Physical entanglement generation rate (network bandwidth) has a massive impact on resource requirements. The entanglement generation rate needs to be matched with the qubit's operational speed. For superconducting qubits, entanglement rate of at least $4$-$5$MHz is required for most applications. For slower qubit platforms such as trapped ion and neutral atom qubits, entanglement generation rate of $5$KHz is sufficient.
    \item \textbf{Node sizing:} For very small nodes, networking overwhelms the resource requirement. As node size increases moderately, networking needs of a node remain constant, allowing all new qubits to be used for computation. Early scientific demonstrations can be performed using nodes with as few as 5000 qubits. Nodes with 25,000 qubits are sufficient to execute small algorithms (such as the Ising model) with only a 1.3$\times$ increase in qubit count and execution time compared to monolithic architectures. For large applications,node sizes of 40,000-60,000 qubits are required to keep qubit and time overheads within 3-8$\times$ of monolithic systems. 
    \item \textbf{System organization:} Qubit organization is application and architecture-dependent, but broadly, 25\% (worst case 65\%) of qubits need to be allocated for entanglement distillation, 6\% (worst case 32\%) for magic state distillation and the remaining for data qubits. Linear connectivity between nodes is sufficient to obtain reasonable resource configurations.

\end{itemize}

\section{Background}

\textbf{Logical qubit:} Physical quantum operations are inherently noisy, generating both random bit‐flip ($X$) and phase‐flip ($Z$) errors.

If left unchecked, these errors rapidly degrade quantum states. 
To mitigate this, quantum error correction (QEC) is used to encode the state of a logical qubit across a set of physical qubits. Common examples are quantum repetition codes, the five qubit code and the surface code.

\textbf{Surface Code:}
The surface code is a popular QEC choice for quantum memory and processing blocks due to its ease of physical implementation~\cite{google2025quantum}. A surface code logical qubit is defined on a 2D grid of physical qubits. 

With a \emph{code distance} $d$, a logical qubit occupies a $d\times d$ patch of physical data qubits, with $(d \times d - 1)$ physical ancilla qubits used for the parity check operations interleaved between them. The code distance determines the ability of the code to detect and correct errors. Higher distances offer better logical error rates at the expense of more physical qubits. The code distance also determines the \emph{cycle time}, denoted by $\tau(d)$. In each cycle, the surface code measures all its parity checks $d$ times to provide sufficient error correction~\cite{Fowler2018}.

\textbf{Logical operations:} The surface code supports single-qubit and multi-qubit operations such as  initialization, measurement and more. Each operation involves one or more rounds of partity check measurements. Therefore, it is common to express operation times in terms of multiples of $\tau(d)$. For example, a logical qubit measurement operation takes 1 cycle ($1\tau(d)$ time). Lattice surgery is the leading technique to implement surface code operations. We assume lattice surgery-based operations~\cite{Fowler2018, Litinski2019}, but details are not important to understand our work. 

\textbf{Multi-Qubit Pauli Measurement (MQPM): } They are a family of surface code operations representing the measurement of joint Pauli observable of multiple qubits. Multi-Qubit Pauli Measurements are implemented \cite{Litinski2019} by initialising an ancilla in the $\bra{0}$ state, entangling it with the state of data qubits involved in the Pauli Observable using CX, CZ or CY gate (depending on the basis in which qubit is measured) implemented by twist-based lattice surgery, and then measuring the ancilla qubit. Depending on the measurement outcome a corrective clifford operation might need to be applied after the measurement.

\textbf{Multi-Qubit Pauli Rotations (MQPR) :} They are a family of surface code operations that are important for large applications and resource estimation (also known as Pauli Gadgets\cite{Cowtan2020} or Pauli Exponentials). They are rotation operators defined by $\exp ({i\alpha \bigotimes_{q\in S}\,P_q})$ where $P_q$ are Pauli operators on a set of qubits S. When $Q$ consists of a single Pauli basis such as X, the Pauli gadget is simply a standard $R_x(\alpha)$ rotation gate. Multi-Qubit Pauli Gadgets are implemented similarly to MQPMs, by applying a rotation gate to an entangled state of several qubits \cite{Beverland2022-kd}. In the case of $\pi / 8$ rotation they can instead be implemented by state injection, applying a MQPM on the involved qubits and a high fidelity magic state \cite{Litinski2019}. Similar to MQPMs, a Clifford correction might be necessary depending on the outcomes of the measurements performed as part of this operation. These operations arise naturally in a number of quantum algorithms (most notably Hamiltonian simulation) and enable effective compilation through commutativity rules \cite{Cowtan2020}.

\textbf{Magic state distillation:} Magic states are a key component of general quantum computation. Logical quantum operations are split into Clifford and non-Clifford operations. Clifford operations can be realised easily in surface code logical qubits and include common instructions such as Controlled NOT and qubit initialisation. Clifford instructions, however, are not sufficient for general quantum computation. Non-Clifford instructions, such as T gate, cannot be executed directly by surface code logical qubits and require a consumption of a special non-Clifford state called magic state $\ket{m} = \frac{1}{\sqrt{2}}(\ket{0} + e^{i\pi/4}\ket{1})$. Magic states prepared directly in hardware are noisy and require being "distilled" by specialised error detection circuits called \textbf{magic state distillation factories (MSDFs)} which consume several low-fidelity magic states and reduce them into fewer higher-fidelity magic states. A common factory is the 15-to-1 factory \cite{LitinskiDistillation2019} which takes 15 low-quality magic states as input and produces 1 high-quality magic state as output. To obtain magic states of arbitrarily high fidelity, multiple factories are chained together, feeding the output of one factory as the input of another.

\textbf{Bell State:} A Bell state is an entangled state of two qubits ($\frac{1}{\sqrt{2}}(\ket{00} + \ket{11})$) that is commonly used for quantum communication and networking.

\textbf{Entanglement Distillation :} 
Networked application operations required high-fidelity Bell states (below $10^{-9}$ error rate), but physical entanglement generation devices can only generate low-fidelity states (typically with error rates above 1\%).
To generate high-fidelity Bell states, we require \textbf{Entanglement Distillation Factories (EDFs)} which, similar to MSDFs, consume several low-fidelity Bell states and use error detection to distil them into a smaller number of higher-fidelity Bell states. Techniques used in EDFs are described in Section \ref{sec:distillation}.

\section{Research Questions and Our Approach}

Our broad research question is ``\textit{What is a feasible architecture for a distributed quantum computer that can offer resource-efficient executions of practical-scale quantum applications?}'' 

This leads to component-level research questions including, ``\textit{How should individual compute nodes be organised in terms of entanglement distillation, magic state distillation and logical qubits?  What node sizes lead to overall resource-efficient executions? What are the costs of entanglement distillation and how should network parameters like entanglement generation rate be selected?}'' To answer these questions, we require new techniques and abstractions at different layers of the quantum stack. 

\begin{figure}[t]
    \centering
    \includegraphics[width=1\linewidth]{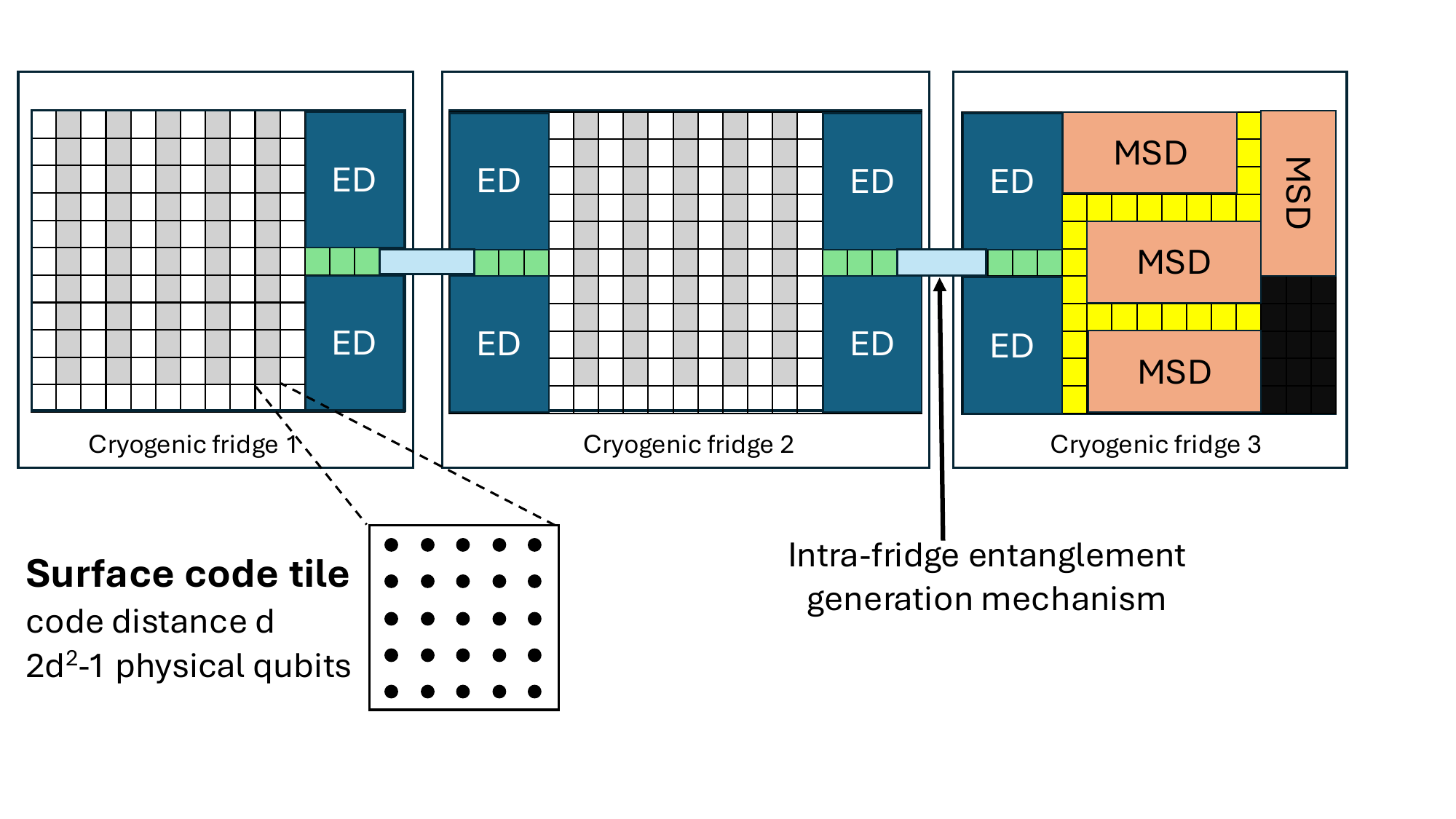}
    \caption{Distributed quantum supercomputer with 3 nodes. Compute nodes store logical data qubits (grey) in surface code tiles arranged inside a fast block layout \cite{Litinski2019}; they are surrounded by ancilla qubits (white) which facilitate operations such as MQPM. Nodes are connected by entanglement generation devices (teal) which generate noisy Bell pairs which are then distilled by Entanglement Distillation (blue) blocks. (Green) space between the ED blocks is used for noisy Bell state routing and distilled Bell state storage. Distillation nodes contain Magic State Distillation (orange) blocks which generate magic states needed for non-Clifford operations; the spaces between them (yellow) are used for distilled magic state storage and network connection (they are used to perform a MQPM between Bell state shared with the node on it's left and node on it's right, thus generating the entanglement between the two nodes). The (black) space represents unused qubits.}
    \label{fig:node-anatomy}
\end{figure}

\textbf{Our overall architecture and ISA:} In Figure \ref{fig:architecture}, we propose an architecture where a number of quantum nodes are connected via a high-performance quantum network. The network connection between neighbouring devices is fulfilled using an entangling device. This device generates shared Bell states, with each neighbour holding one physical qubit of the Bell state. 
As shown in Figure \ref{fig:node-anatomy},  we consider a linear network topology, with each node connected to at most two other nodes. This allows us to minimise networking overhead while not significantly affecting the runtime of computation (as discussed in Section \ref{our_architecture}).

At the lowest layer, each node consists of physical qubits which are used both as compute qubits and for network interfacing. On top of this layer, compute qubits are used to implement surface code-based logical qubits and MSDFs.

The network interface qubits are used to store distributed Bell states and to implement EDFs. Together, these components are used to implement \emph{distributed planar quantum ISA} which adds network operations to an existing quantum ISA~\cite{Beverland2022-kd, Litinski2019}.

\textbf{Compilation technique:}
We develop a compilation technique that is able to transform algorithms designed for monolithic quantum computers into distributed planar quantum ISA instructions. We take MQPR based compilation developed by Litinski et. al.\cite{Litinski2019} as the baseline and add a technique for implementing multi-node MQPRs in a distributed setting. Our technique uses local Multi-Qubit Pauli measurements as an entangling operation, and Bell States as a way of sharing entanglement between nodes. This technique allows us to model the execution of general applications on a distributed system.

\textbf{Modelling entanglement distillation:} 

Entanglement distillation is the heart of distributed quantum computation, but accurate physical noise models are lacking in literature. We develop a modelling tool which carefully simulates the error correction performance of a range of entanglement distillation units on noisy hardware.  Our tool relies on error enumeration techniques developed in \cite{Beverland2021, LitinskiDistillation2019}. We then consider different ways of combining these units into multi-level EDFs, giving us access to Bell states with high fidelities that are necessary for applications. Using this tool we are able to consider a large space of factories with different output error rate, space, time and input efficiency trade-offs.

\textbf{Distributed quantum resource estimator:} To combine all the elements above, we develop a new resource estimation tool. Taking compiled application operation and qubit counts, architecture models and hardware parameters as input, this tool estimates the resource requirements of algorithms on our proposed architecture and outputs node count and execution time. For an application, we have the choice of selecting the type and configuration of MSDFs and EDFs as well as the surface code distance of the compute block. To minimise resources, we consider the space of different factories that fulfil the application's error budget (desired accuracy) and their counts. For each such configuration, we calculate the application runtime (gated independently by surface code cycle time, high fidelity magic state production rate across the network and high fidelity Bell state production rate at every interconnect), required surface code distance and number of nodes required to host all the qubits. Out of these configurations we pick ones with best node count - runtime trade-off to give us the final estimate of application resource requirements.

\section{Compilation Process}
\label{sec:compiler}

First, starting from a general quantum application written in terms of quantum gates and measurements, a Clifford optimisation pass developed by Littinski \cite{Litinski2019} is applied. This removes Clifford operations entirely from the circuit by converting every quantum gate into a Multi-Qubit Pauli Rotation representation and, using commutativity rules, moves all Clifford MQPRs to the end of the circuit, absorbing them into measurements. This results in a dramatically reduced instruction counts and program depth. This is a standard technique employed by current resource estimators and architecture simulations \cite{Beverland2022-kd, xqsim, harrigan2024qualtran}. Further, it allows logical qubit operations to be implemented without moving data qubits, reducing the need for routing passes which are commonly required in other quantum compilation tasks. As a result of this pass, the application is transformed into a form where it consists of a series of MQPRs with $\alpha=\pi/8$ (non-Clifford) and Multi-Qubit Pauli Measurements. 

In a monolithic scenario, $\pi/8$ MQPRs are executed by preparing a T-state and performing a Multi-Qubit Pauli Measurement of involved qubits and the T-state.
The ancilla used for the T-state is then measured, and the combined result of these two measurements is used to determine a Clifford correction operation that needs to be applied. These corrections do not change resource estimates and can be ignored (the Clifford correction gets commuted to the end of the circuit, getting absorbed into measurement, and changing the bases of subsequent MQPRs in software).

We introduce a novel way of decomposing multi-node MQPRs on distributed quantum computers. 
At a high-level, we achieve this by separating the computation into a set of local MQPM operations and composing them using shared Bell states across devices. This is similar to how computations such as matrix operations are expressed in classical supercomputers as local operations followed by communication operations, except in quantum setting communication is realised through Bell state measurement.  

The transformation is demonstrated in Figure \ref{fig:nnodepauli} and proof of it's correctness can be found in the Appendix. Specifically, each pair of adjacent nodes (in a linear connectivity) first shares a high fidelity Bell pair and a magic state is prepared on one of the nodes. A MQPM is then performed on each node, involving corresponding data qubits, Bell states and (on the node which holds it) the magic state. It is important that even if the node does not contain any data qubits, it still performs MQPM in order to maintain the entanglement of the nodes to it's left and it's right. Finally, the ancilla qubits holding Bell states and magic state are destructively measured. The combined result of all the MQPR and ancilla measurement are  used to determine what Clifford correction is required. The correction MQPRs can be commuted through the circuit and absorbed into the measurements algorithmically at runtime, just like was done during the compilation.

Post-decomposition, the program consists of layers of operations, each layer representing the execution of a single MQPR or MQPM. A layer consists of 
magic state  and Bell state initialisations, followed by local MQPMs and measurements of ancillae that used to hold the magic state and Bell states. Finally, the results of these measurements are consumed by the runtime environment in order to determine what correction needs to be applied to the subsequent measurements. 


\begin{figure}[t]
    \centering
    \includegraphics[width=1\linewidth]{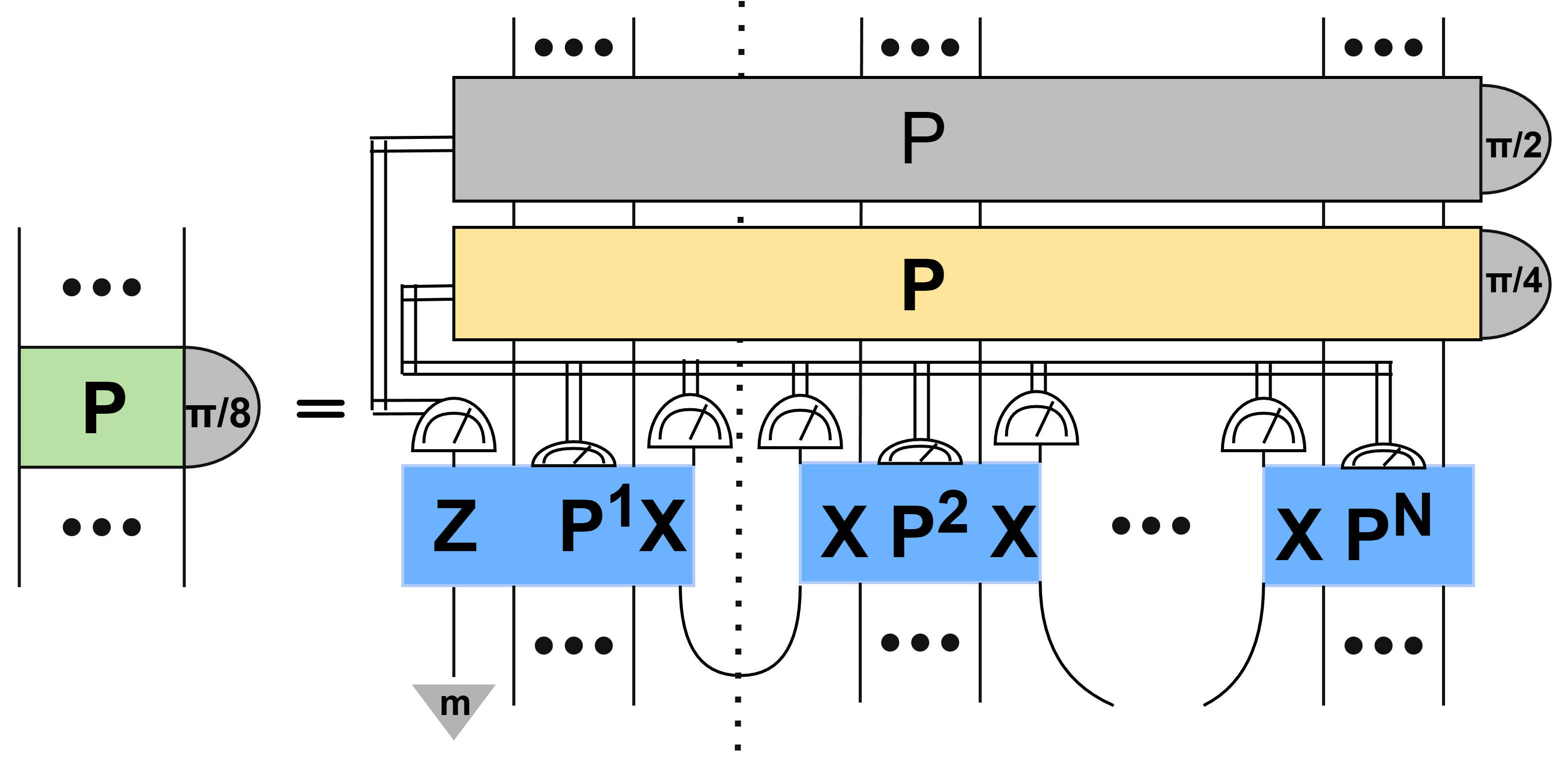}
    \caption{Our remote implementation of MQPR. Here, time flows from bottom up and single lines represent qubit "wires". LHS is the MQPR with angle $\pi/8$ and Pauli string $P = P_1P_2...P_n$ performed on the monolithic device as described in \cite{Litinski2019}. We assume that data qubits represented by the Pauli string $P_i$ all reside on the same node $N_i$, separated by dotted line in the diagram. RHS represents the operation performed on a distributed quantum computer utilising only local operations, bell states and classical processing. Magic state is represented by the grey triangle while bell states are represented with bent wires.  Blue boxes represent local MQPMs where data qubits are measured in basis corresponding to the pauli string $P_i$, magic state is measured in Z basis and bell states are measured in X basis. Doubled wires represent classical signal from the measurement which is processe by the control computer to determine whether correction operations (Grey and Yellow box) are necessary. These operations can be performed purely classically with Pauli frame tracking and runtime adjustments to further operations. We prove the equivalence of LHS and RHS in the appendix.}
    \label{fig:nnodepauli}
\end{figure}

\section{Our Overall Architecture and ISA}
\label{our_architecture}

\subsection{Network}

Building a quantum interconnect between two nodes requires shared entanglement using Bell pairs(a loose analog in classical networking is the establishment of links in a virtual circuit network). Once Bell pairs are shared between the nodes there are two mechanisms by which they can be consumed: 1) using raw Bell pairs generated by the interconnect to replicate physical operations, allowing noisy operations between the physical qubits on different nodes \cite{Ramette2024} or, 2) distilling raw Bell states using entanglement distillation \cite{Dür_Briegel_2007} into high-fidelity logical Bell pairs which are then consumed as a resource with logical operations. Which approach is suitable for scale? 

It has been demonstrated that if the error rate of the remote entangling operation is within an order of magnitude of the error rate of local operations, then the raw Bell pairs can be used directly for physical operations in order to perform lattice surgery between multiple nodes without significantly contributing to the total error budget \cite{Ramette2024}. This result, however, is inapplicable in the foreseeable future of the Fault Tolerant superconducting computing. Firstly, to utilise remote operations for lattice surgery, every qubit along the edge of the surface code patch needs to have the capability of remote operation, since movement of noisy unencoded Bell states would contribute significantly to the fidelity of Bell state, taking it outside the noise threshold. This significantly complicates the architecture, as entangelment generation devices would have to be miniaturised to the scale of individual physical qubits. Second, when using remote operations for lattice surgery directly, the speed mismatch between remote operations and local operations has higher impact on computation runtime, as remote operation frequency is directly linked to the frequency with which operations can be executed. Lastly, this result relies on remote operation fidelity being within an order of magnitude of local operations; this high of a fidelity has not yet been demonstrated experimentally and there is no reason to believe that this could be achieved in the foreseeable future while maintaining the requisite frequency.

Hence, we believe that studying the effects of entanglement distillation on resource requirements of medium-scale distributed quantum computers is essential to accurate modelling. Entanglement distillation alleviates all the issues highlighted above - it allows the architecture to handle higher interconnect error rates, it allows the architecture to compensate for slower network frequency by using error detection codes with higher input/output ratio at the expense of utilising more local on-chip resources and it allows the interconnects to be spaced wider apart on the chip as the Bell states can be encoded into surface code patches immediately after generation and the transit errors can be tolerated more easily.

We assume a linear network topology among the $N$ nodes. Prior works such as \cite{Feng2024} theoretically studied distributed quantum algorithms assuming a star network topology.  For most algorithms, our results show that entanglement distillation for a single network connection requires on average 12\% of node's qubits, reaching as high as 32\% in the worst case. Therefore, it is costly to maintain excess connections, both on a network and on a node level. Increasing the number of connections with a limited qubit budget also forces us to reduce the number of EDFs per connection, which can slow down the computation. Limiting the network to minimal linear connectivity decreases the overheads incurred by networking and is sufficient for sequential remote MQPR execution as demonstrated in the appendix. Furthermore, linear connectivity in which each node is connected to at most two neighbours splits the load of entanglement distillation evenly between the nodes.

\subsection{Node Architecture}

At the architectural level every node needs to be able to generate entanglement with other nodes, store logical data qubits and perform MQPMs on any subset of those data qubits. Some of the nodes need to also be able to generate high quality Magic States to be consumed through state injection in order to perform $\pi/8$ MQPRs.

We assume that we have a network of N nodes. Each node in the distributed system is a collection of data qubits, ancilla qubits used to perform operations, MSDFs and EDFs. At this level, we can abstract away certain physical qubit details such as physical qubit layout or gate times. The logical qubits are implemented using the surface code, with a distance $d$ that is set based on application and hardware parameters. MSDFs and ESDs also use surface codes under the hood, but they can typically use a lower code distance than the logical qubits used for mapping algorithms.

Within each compute node, we require the ability to perform MQPMs on an arbitrary subset of data qubits to effectively support large applications. The easiest way to realize this is to leverage Litinski's \cite{Litinski2019} monolithic "fast compute block" architecture within a node. Here, $Q_D$ data logical qubits are surrounded by $Q_D + \ceil{\sqrt{8Q_D}} +1$ ancilla logical qubits which enable MQPMs. Fast block encoding efficiency (number of logical qubits stored per physical qubit) increases with size. Hence, we opt to position all MSDFs on dedicated distillation nodes, increasing the encoding efficiency of compute nodes.

Distillation nodes, in turn, consist of MSDFs and routing lanes. MSDFs produce high fidelity magic states which are then consumed as a resource to implement MQPRs

In addition, for each node's network connection, we provision EDFs within the node to distil Bell states generated by the connection. This architecture is showcased in Figure \ref{fig:node-anatomy}.

\subsection{Planar Distributed Quantum ISA}

To model resource needs, we need a quantum instruction set architecture (ISA) for a distributed system. This allows us to concretely model the instruction counts and logical cycles for application. Prior works developed the Planar Quantum ISA for monolithic systems~\cite{Beverland2022-kd, Litinski2019}. We add network operations to this ISA.  We  assume that computation happens on a grid of tiles, each tile corresponding to a single logical qubit encoded in the surface code. Below, the surface code's cycle time  is denoted as $\tau (d)$, physical initialise time as $t_{init}$ and physical measurement time as $t_{meas}$. Below we list out the full ISA for completeness, our contribution is the last instruction:
\begin{itemize}
    \item \textbf{Initialisation} of either a 1-tile patch storing the state of one qubit in the $\ket{0}$ state or a 2-tile patch storing the state of 2 qubits in $\ket{00}$ state with cost $0 \tau (d) + t_{init}$.  
    \item \textbf{Pauli} gates applied to qubits stored in 1 or 2-tile patches with time cost $0$ executed by tracking the Pauli frame change in software~\cite{PauliFrameTracking, Riesebos2017}.
    \item \textbf{Hadamard} or \textbf{Phase} gates applied to qubits stored in a 1 qubit patch with cost $3\tau(d)$ and $2\tau(d)$ respectively. Both of these operations require an adjacent ancilla, which is provided by the architecture.
    \item \textbf{Destructive measurement} of qubits in a one-tile patch or two-tile patch with cost $0\tau(d) + t_{meas}$
    \item \textbf{Movement} of a qubit in a one-tile patch into another one-tile patch connected by a continuous ancilla region.
    \item \textbf{Non-destructive multi-qubit Pauli measurement (MQPM)} of any number of qubits in 1-tile or 2-tile patches can be performed with cost $1\tau(d)$. This operation measures XOR of involved qubits in different bases. The operation is performed by initialising a large ancilla region neighbouring all involved qubits and measuring the boundary between the ancilla and each qubit. The ancilla tiles needed for this are drawn from the $Q_A+\ceil{\sqrt{8Q_A}}+1$ tiles provisioned in Section \ref{our_architecture}.
    \item \textbf{T-state preparation} in a one-tile patch neighbouring a MSDF performed by using the MSDF to distil the state. The time cost of this operation depends on the MSDF being utilised, typically between $11 \tau (d)$ and $24 \tau (d)$ with every factory being able to operate in parallel.
    \item \textbf{Bell-state preparation} of two qubits on two connected nodes. A half of Bell state appears on the node in a one-tile patch neighbouring an EDF used to distill the state. The time cost of this operation depends on the EDF being utilised, typically between $3 \tau (d)$ and $17 \tau (d)$.
\end{itemize}

During the algorithm runtime we assume that EDFs and MSDFs are running continuously. MSDFs generate magic states which are stored in the corresponding node's memory. EDFs generate Bell states which are stored in the interconnect's memory. In order to perform intra-node MQPR, one Bell state per network connection and one magic state in total need to be present. Once these conditions are fulfilled, each  node performs a local MQPM involving a Bell state for each connection, all requisite data qubits and exactly one magic state. Note that even if no data qubits of a particular node are involved in a particular intra-node MQPR, it still needs to perform local MQPM between Bell states to maintain the network connectivity.

The "clock time" of the system (rate at which MQPRs can be implemented) is independently gated by Magic State generation rate, every interconnect's Bell State generation rate and the time it takes local data blocks to execute a local MQPM. Balancing these three variables to achieve optimal space-time performance is the main task of resource estimation. Our tool considers a wide space of potential EDFs, MSDFs and their numbers to provide the best match between the speeds of these components while also considering their space footprint on the device.

\section{Modelling Entanglement Distillation}\label{sec:distillation}

Unlike MSDFs which rely on QEC codes that must support transversal T gats, EDFs can employ a higher variety of codes. To our knowledge, there are no existing frameworks for modelling the performance of broad range of EDFs on noisy systems. We created an entanglement distillation framework for use as a part of our resource estimation pipeline. It allows us to model a broad range of factories, including new MSDFs.

\subsection{Entanglement Distillation Factories}

We can transform any QEC with a known efficient decoder  into an EDF~\cite{Dür_Briegel_2007,ediss3588}. We start with a set of Bell pairs shared between two nodes. To perform error detection, the decoding circuit is applied to both sides of the Bell pairs and the axillary qubits of the code are measured, giving us an "error syndrome". If the error syndromes differ between the two sides, an error is detected and the result is discarded. Otherwise, a correction gate is applied to one side based on the syndrome. A larger number of codes with no known efficient decoding circuit can be utilised using a technique described in \cite{PRXQuantum.6.010339}.

For this paper we chose to consider two different codes with an efficient decoding circuit. This small selection of codes allows us to consider a large number of combinations for multi-level distillation factories. 

\noindent\textbf{2-qubit repetition code} is the code obtained by encoding a state of the qubit by "duplicating it". The decoding circuit simply measures the entangled state of these two qubits in a particular basis. This yields Z, X and Y basis versions of the code which can detect bit flip, phase flip or combinations of bit and phase flip errors . It is the smallest possible QEC and can be efficiently implemented on top of surface code logical qubits, requiring only 2 logical qubits, with time cost $2\tau(d)$. A distinguishing feature of this code is that it produces output with infidelity skewed heavily towards one basis of measurement, meaning that in order to efficiently utilise it in a multi-level factory setup X, Y and Z basis errors need to be considered separately.

\noindent\textbf{5-qubit "perfect" code} is the smallest code with perfect theoretical information efficiency \cite{Bennett1996, GVBound} .
Because of this, it demonstrates a good input efficiency, outperforming repetition codes when heavily constrained by entanglement generation rate. Unlike repetition code, this code required a number of axillary logical qubits in a 2D surface code architecture; we found a way to implement the code with 15 logical qubits and time $3\tau(d)$.

We model EDFs by enumerating the errors arising from input Bell states and from Clifford operations of the decoding circuit following similar methodology for magic state error modelling as~\cite{LitinskiDistillation2019, Beverland2021} . We separately consider X, Z and Y basis errors and simulate the entire circuit using density matrix representation with each individual error present. For input errors we consider all the error combinations and for Clifford errors we assume that all three kinds of errors have equal probability and only consider one error happening at a time. The resulting models are shown in Table \ref{tab:Bellfact}.

 \begin{table*}
 \small
     \centering
          \caption{Entanglement distillation error models computed by our work. Models are for 5Q (5-qubit perfect code) and 2Q (2-qubit reptition code with different basis) factories. Probability of error being detected and as such factory output being rejected or an undetected error occuring in terms of X, Y and Z input Bell state error rate ($P_{X/Y/Z}$) and physical error rate $p$ }
     \scalebox{0.8}{
     \begin{tabular}{ccccc}
         Factory & Rejection Probability & X error Probability & Y error Probability & Z error Probability\\
         \hline 
         5Q & $5P_{Z}+5P_{Y}+5P_{X}+6.5p$ & $5P_{X}P_{Z}^2+5P_{X}P_{Y}^2+3.1p$ & $5P_{Y}P_{Z}^2+5P_{Y}P_{X}^2++1.7p$ & $5P_{Z}P_{Y}^2+5P_{Z}P_{X}^2++1.7p$\\
         2Q(X)  & $2P_{Z}+2P_{Y}+2P_{X}P_{Z}+2P_{X}P_{Y} +2.4p$&  $2P_{X}+0.8p$&  $2P_{Y}P_{Z}+0.8p$&  $P_{Y}^2+P_{Z}^2+0.8p$ \\
         2Q(Y) & $2P_{Z}+2P_{X}+2P_{Y}P_{X}+2P_{Y}P_{Z} +2.4p$&  $2P_{X}P_{Z}+0.8p$&  $2P_{Y}+0.8p$&  $P_{X}^2+P_{Z}^2+0.8p$ \\
         2Q(Z) & $2P_{X}+2P_{Y}+2P_{X}P_{Z}+2P_{Y}P_{Z} +2.4p$&  $P_{Y}^2 + P_{X}^2+0.8p$&  $2P_{Y}P_{X}+0.8p$&  $2P_{z}+0.8p$ \\
     \end{tabular}
    }
     \label{tab:Bellfact}
 \end{table*}

\subsection{Multi-level factories}
To arbitrarily improve the fidelity of output Bell states we chain several factories together, feeding the output of one set of factories as an input to another. Note that all codes considered by us only produce one output, allowing us to ignore the effect of correlated errors. We start with physical Bell states produced by the entangling device, treating them as an output of level 0. We then repeatedly use outputs of level $n$ factories as inputs of level $n+1$ factories. As the fidelity of the factory's output increases, errors introduced by Clifford operations in the decoding circuit become the major contributor to the output error. To mitigate this, we run every layer of factories on an underlying surface code. In order for the Clifford operation fidelity to keep up with the fidelity of the output Bell states, between every two layers of factories we may need to increase the surface code distance. Further, to satisfy the input needs at level $n+1$, we require multiple copies of factories at level $n$. These configuration decisions of multi-level factories are handled by our tool.

These factories are characterised by their qubit footprint, runtime, number of inputs/outputs and error rate of the distilled states. We evaluate a large variety of multi-level factories to find ones with best combinations of parameters. An example of one such factory is 6 copies of repetition code (Z) with distance 1 $\rightarrow$ 1 copy of 5-to-1 code with distance 3 $\rightarrow$ 1 copy of repetition code (Z) with distance 7 $\rightarrow$ one copy of repetition code (X) with distance 9 which occupies 324 qubits for 873 time steps, requires 56 inputs (on average) and produces a Bell state with error probability of $4.5 \times 10^{-10}$. In general, our experiments showed that multi-level factories consisting of different basis versions of the repetition code consistently outperformed all the other factory setups in terms of space-time efficiency, at the cost of low input efficiency (requiring more inputs for similar error rate). 

\section{Distributed Quantum Resource Estimator}
\textbf{From Applications to ISA-level estimates:} Our framework takes a quantum algorithm written in a high language and error budget $\epsilon$ (1 - desired precision) as input. Error Budget is split into three parts, $\epsilon = \epsilon _L + \epsilon _M + \epsilon _E$, accounting for total error arising from logical qubits, magic states and Bell states. This additive error model is standard in resource estimation studies and arises from a union bound analysis of application failure modes~\cite{Beverland2021, Litinski2019}. We apply a monolithic resource estimation tool such as Azure RE \cite{Beverland2022-kd} to determine the $Q_D$, the number of data qubits and $T_{count}$, the T-gate count after the MQPR transformation described in Section \ref{sec:compiler}. At this stage, the program consists of $T_{count}$ sequential MQPR operations, each performed in a manner described in Section \ref{sec:compiler}.


\noindent\textbf{From ISA-estimates to physical estimates:} To obtain physical estimates, we choose the appropriate  factories. Multi-level MSDFs and EDFs are pre-computed by our tool for a given physical qubit and network error model. Each such MSDF (EDF) is parametrised by qubit count  $Q_M$($Q_E$), runtime $Q_M$($Q_E$), number of input states $I_M$($I_E$), number of output state $O_M$ and output error rate $\epsilon _M$($\epsilon _E$). Out of all available multi-level MSDFs and EDFs, we choose the ones that meet the precision needs of the application. The overall contribution of magic and Bell states to the application error is $T_{count}(\epsilon_{M} + \epsilon_{E}(N-1))$, which must be lower than $\epsilon _M + \epsilon _E$. For each MSDF and EDF we determine the possible factory counts ($N_M$, $N_E$) which depend on the node size and Bell state generation rate $\eta$. 

For each factory configuration we then jointly determine the surface code distance $d$ for the compute blocks and number of nodes needed to host all the data qubits. If the node size is $Q_{Node}$, then each "edge" node has $Q_{Node} - N_EQ_E$ physical qubits available and each "central" node has $Q_{Node} - 2N_EQ_E$. These physical qubits are split into logical tiles, each requiring $2d^2-1$ qubits; the tiles are arranged in the largest fast block layout possible with the available number of tiles, producing $Q_{central}(Q_{edge})$ - the number of logical data qubits stored in the central/edge node respectively.

Similarly, we determine the required number of magic state distillation nodes, where each such node fits $(Q_{Node} - 2N_EQ_E)/Q_M$ factories. Edge cases where leftover data qubits / factories can be fitted into a single node are considered separately/

For each configuration we determine how often a MQPR can be executed, which is constrained by the surface code cycle, runtime of factories and $\eta$:

\[T_{MQPR} = \text{max} \left( \tau (d), \dfrac{T_M}{N_M \times O_M}, \dfrac{T_E} {N_E \times O_E}, \dfrac{N_E \times I_E}{ \eta}\right)\]

Since we model a quantum supercomputer where all nodes are within the same datacenter, we consider a control architecture where classical control or decoding latency are minimal and a second-order effect in comparison to the logical cycle time of the quantum processor and Bell state generation time (tens of microseconds) ~\cite{harrigan2024qualtran,Beverland2022-kd}

We use this number to get the runtime of computation $T_{MQPR} * N_T$. Given runtime, we check that the surface code distance $d$ is high enough to store data qubits and ancillae for the entire runtime duration while maintaining error below the budget:
\[\epsilon(d) \le \dfrac {\epsilon _L} {Q_L \times T_{count} \times (T_{MQOR}/\tau(d))}\]

Where $\epsilon (d)$ (logical error rate of surface code with distance $d$) as a function of physical error rate $E_{phys}$ is $\epsilon(d) = 0.03 \times \left(\dfrac{E_{phys}}{0.01}\right) ^ { (d+1)/2}$ \cite{Fowler2018}

If the chosen surface code distance is insufficient, we backtrack and choose higher distance.

\section{Experimental Setup}
\begin{table}[t]
  \caption{Benchmark applications. These applications are relevant for quantum advantage demonstrations. Parameters were obtained using the Azure Quantum Resource Estimator~\cite{Beverland2022-kd} and Qualtran~\cite{harrigan2024qualtran}.}
\small
  \centering
  \begin{tabular}{llrr}
    \toprule
    \textbf{Category} & \textbf{Application} & \textbf{Qubits} & \textbf{$T$‑gates} \\
    \midrule
    \multirow{3}{*}{\shortstack{Quantum\\Dynamics}}
      & Ising $10\times10$                  & 100    & $9.54\times10^5$          \\
      & Fermi–Hubbard $10\times10$          & 241    & $7.93\times10^8$          \\
      & Heisenberg $10\times10$             & 123    & $2.55\times10^{10}$       \\
    \midrule
    \multirow{3}{*}{\shortstack{Molecular\\QPE}}
      & ZnS QPE             & 351   & $6.12\times10^{10}$       \\
      & Benzene (C$_6$H$_6$) QPE   & 504   & $2.86\times10^{11}$       \\
      & Ruthenium complex QPE               & 1,318   & $2.70\times10^{11}$       \\
      & Nitrogenase cluster QPE             & 1,424   & $8.63\times10^{12}$       \\
    \midrule
    Factoring
      & 2048‑bit RSA integer                & 12,581  & $1.50\times10^{10}$       \\
    \bottomrule
  \end{tabular}

  \label{tab:applications}
\end{table}


\textbf{Applications:} We consider three classes of quantum algorithms: many-body dynamics simulation~\cite{hatano2005finding}, quantum phase estimation for ground-state energy calculation~\cite{babbush2018encoding}, and Shor’s integer factoring~\cite{gidney2021factor}. Table \ref{tab:applications} lists their problem sizes. We do not consider NISQ benchmarks such as VQE or QAOA as their quantum advantage and scaling prosects are  unclear and they are design to evaluate NISQ hardware, rather than FTQC systems. Due to our use of Clifford optimization from \cite{Litinski2019}, T-gate count is a good proxy for runtime. 
The details of these applications can be found in the appendix. We emphasize that application-level benchmarking is inherently coupled to algorithmic choices, and resource estimates can be substantially improved by optimizing the underlying quantum algorithms. For a fixed physical task, such as quantum dynamics simulation, switching between algorithmic paradigms (e.g., Trotterization versus qubitization) can lead to qualitatively different space-time overheads, gate counts, and magic-state consumption. As a result, improvements observed in resource estimates may originate from different layers of the stack, including algorithm design, compilation strategies, or architectural assumptions, rather than from the hardware or error correction layer alone. Care must therefore be taken when comparing benchmarks to ensure fairness and interpretability, particularly when the optimal algorithmic realization for a given application is not yet known.

\textbf{Hardware parameters:} We model all physical operations take the same amount of time $t_{op}$, which is set to the runtime of longest operation. We mainly focus on superconducting qubit architectures, setting $t_{op} = 50 ns$ ~\cite{Beverland2021}, but we also run a set of early quantum advantage experiments with slower clock speed $t_{op} = 100 \mu s$ ~\cite{Beverland2021} which is more representative of neutral atom and trapped ion architectures. We explore both a pessimistic physical error rate of $.1\%$ and an optimistic error rate of $.01\%$ based on hardware projections~\cite{Beverland2021}. 
For entanglement error rate we consider three scenarios: a current value of $5\%$ ~\cite{Main2025}, optimistic value of $1\%$ and DARPA target of $0.1\%~$\cite{DARPA}.
For entanglement rate, we consider a range, 300Hz-200MHz to inform hardware targets.
We consider node sizes between 3000 qubits and 100000 qubits, covering a range of possibilities from near-term to very-long term devices.

\textbf{Metrics:} For each application and architecture (QEC, MSDF, EDF, physical qubit, network parameters), our estimator produces the total number of physical qubits required and the application runtime.

We compare different architectures in terms of space-time volume (number of qubits $\times$ runtime) of applications being executed on them. 
For every architecture point there are multiple factory configurations which fulfil the application needs, offering different trade-offs between runtime and number of qubits (example in Table \ref{tab:early-ising}). 
We select the configuration with the lowest spacetime volume as a representative rather than targeting space-optimal or runtime-optimal configurations.
To compare the cost of distribution of different application against each other, we normalize by the space-time volume of the corresponding application being executed in a monolithic setting, i.e. in terms of overhead incurred by distributing the application. \textbf{Lower normalized space-time overhead is better.}

Our toolflow is written in Python 3.12.8 and we use resource counting functionality of Azure Quantum Resource Estimator from ~\cite{AzureQuantum}.  
\section{Results}\label{sec:results}

\subsection{Validation of our Toolflow}
Table \ref{tab:estimatecomparison} compares monolithic resource estimates produced by our tool with the Azure Quantum Resource Estimator. Our estimates match or outperform Azure RE for most applications. This is because our estimator includes a 20-to-4 MSDF, which we were able to model as a side result of the framework we developed for EDF modelling.  These results give us confidence that our estimates for the local parts of distributed quantum computation are accurate.  

\begin{table*}[ht]
\small
  \caption{Comparison of resource estimates produced by Azure Resource Estimator and our tool. All estimates assume $50ns$ qubits with error rate of $10^{-4}$. For distributed setting, estimates assume $10MHz$ entanglement generation rate, either 1\% or 0.1\% Bell state error rate and node size of $45,000$ qubits.}
  \centering
  \scalebox{0.7}{
  \begin{tabular}{lrlrlrlrlrlr}
    \toprule
    \multirow{2}{*}{\textbf{Application}} &
      \multicolumn{2}{c}{\textbf{Monolithic - Azure}} &
      \multicolumn{2}{c}{\textbf{Monolithic - Ours}} &
      \multicolumn{2}{c}{\textbf{Distributed - 5\%}} &
      \multicolumn{2}{c}{\textbf{Distributed - 1\%}} &
      \multicolumn{2}{c}{\textbf{Distributed - 0.1\%}} \\
                          & Qubits  & Runtime    & Qubits  & Runtime    & Qubits  & Runtime     & Qubits  & Runtime     & Qubits & Runtime\\
    \midrule
    Ising 10x10           & 0.11M & 7.9 sec   & 0.092M & 7.9 sec   & 0.088M & 19 sec    & 0.088M & 14 sec    & 0.12M & 8.0 sec \\ 
    Fermi-Hubbard 10x10   & 0.23M  & 52 min   & 0.26M  & 52 min   & 0.29M  & 3.2 hours   & 0.44M  & 1.6 hr     & 0.44M & 1.6 hr \\ 
    Heisenberg 10x10      & 0.18M  & 1.3 days  & 0.24M  & 1.3 days  & 0.22M  & 4.8 days   & 0.35M  & 2.7 days   & 0.21M & 4.0 days \\ 
    \midrule
    Shor's Factoring 2048 & 12M   & 19 hr    & 8.7M   & 16 hours & 19M   & 2.9 days   & 20M   & 1.3 days   & 20M  & 1.3 days \\ 
    \midrule
    ZnS QPE               & 0.37M  & 3.2 days  & 0.45M  & 3.2 days  & 0.64M  & 13 days   & 0.52M  & 14 days   & 0.53M & 9.7 days \\ 
    Benzene QPE           & 0.89M  & 17 days  & 0.60M  & 18 days  & 1.2M   & 2.0 months & 2.4M   & 30 days   & 0.98M & 2.1 months \\ 
    Ruthenium QPE         & 1.9M   & 16 days  & 1.7M   & 16 days  & 3.6M   & 2.1 months & 2.4M   & 3.0 months & 2.5M  & 2.0 months \\ 
    Nitrogenase QPE       & 2.4M   & 1.6 years & 2.3M   & 1.6 years & 5.1M   & 5.6 years  & 5.1M   & 5.5 years  & 3.6M  & 5.1 years \\ 
    \bottomrule
  \end{tabular}
  }
  \label{tab:estimatecomparison}
\end{table*}

\begin{figure*}[ht]
    \centering
    \includegraphics[width=1\linewidth]{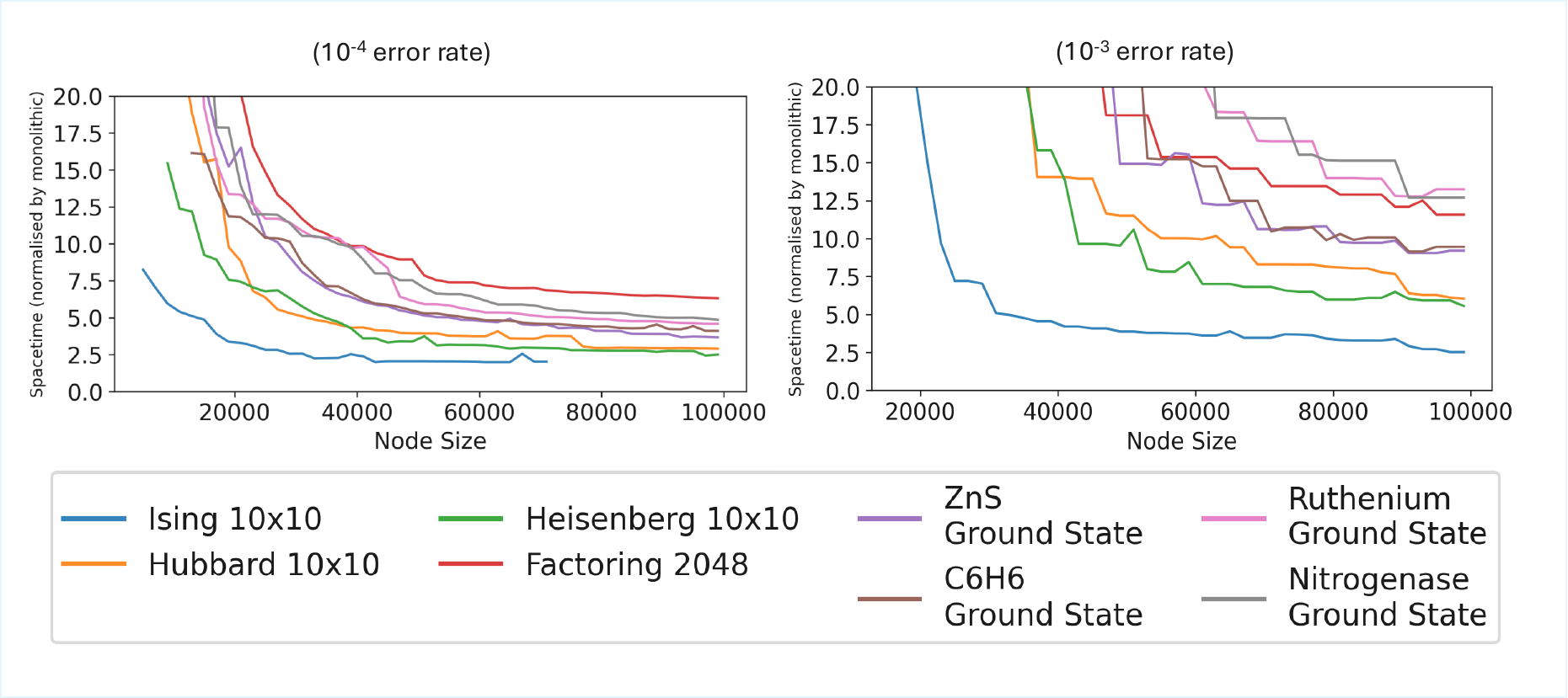}
    \caption{Effects of the node size on the spacetime volume overhead of the distributed computation compared to the monolithic architecture, for 2 different error rates. In every configuration the raw Bell state error rate is $5\%$ and the entanglement rate is $10MHz$. The data shows that in the case of $10^{-4}$ error rate node sizing of 40-60K qubits is ideal across applications and for $10^{-3}$ error rate node sizing of 70-90k qubits is preferred.}
    \label{fig:nodesize}
\end{figure*}

\begin{figure*}[ht]
    \centering
    \includegraphics[width=1\linewidth]{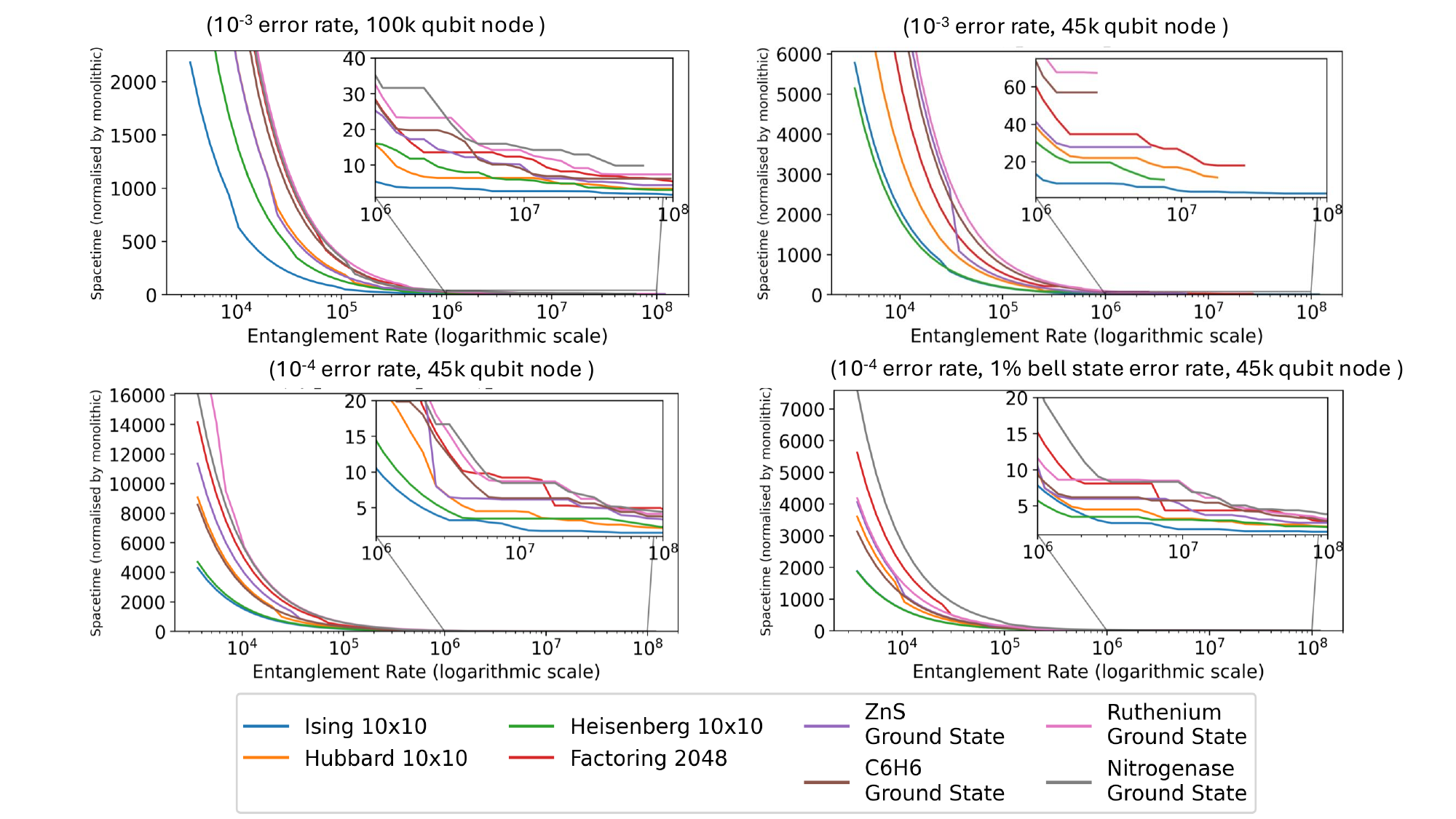}
    \caption{Effect of entanglement generation rate on spacetime overhead of distributed computation compared to monolithic, for 4 different (physical error rate, node size, Bell state infidelity) hardware configurations. Bell state infidelity is $5\%$ for all graphs except bottom right, for which infidelity is $1\%$.}
    \label{fig:entanglementrate}
\end{figure*}

\subsection{Importance of Node Sizing}

Figure~\ref{fig:nodesize} illustrates how node size affects the spacetime overhead in distributed quantum computing. With $10^{-4}$ error rate, node sizes in the range of $40$K-$60$K qubits offer an optimal balance of low space-time overhead and hardware feasibility. When nodes have less than $40$K qubits, we observe very large space-time overheads (typically more than 10X) compared to monolithic systems. This trend is because networking requirements do not scale with the size of the node; smaller nodes need to dedicate an increasingly large proportion of their qubits for network. 

Beyond $60$K qubits, we observe diminishing returns in space-time overhead. 
This result contrasts with IBM’s current roadmap \cite{IBM} which targets $25$K qubit nodes operating on an underlying QLDPC code \cite{GrossCode}. Our work shows that $25$k qubit nodes with underlying surface code architecture are sufficient to run early scientific applications, but could incur over an order-of-magnitude increase in space–time overhead for most workloads as systems evolve. 

\subsection{Matching Entanglement Rate and Qubit Speed}
Figure \ref{fig:entanglementrate} shows the effect of entanglement generation rate $\eta$ on the space-time overhead. 
As $\eta$ increases, the overheads of distributed computing reduce. The graphs exhibit a piecewise linear behaviour. This is because for any given EDF configuration the runtime of the algorithm scales linearly with $\eta$, up to the point where the factories are completely saturated. That is, the EDF stops being bottlenecked by the availability of input Bell states. Once saturated, the runtime for a given factory remains constant no matter how high $\eta$ grows. The step-like shape of the graph arises from our resource estimator picking more space-time efficient (and hence less entanglement efficient) factories as $\eta$ increases. That is, when Bell states are in limited supply, EDFs must make efficient use of scarce Bell pairs. However, when an abundant supply of Bell pairs is available, protocols with low space-time overheads can be used, as the cost of consuming additional pairs is negligible.

Comparing trends across fast and slow qubits, we see that entanglement distillation rates must be matched to hardware speed. Distributed quantum computing is much more easily achievable on slow (neutral atom and trapped ion) qubits in the short term. Even a modest $4$    KHz entanglement generation rate is sufficient to produce very low overheads (table \ref{tab:early-ising}). Such architectures might be useful for early applications such as the Ising model which have short runtimes.  For fast qubits, a much higher entanglement generation rate is required --- for superconducting qubits, at least 4MHz will be required.

\subsection{Informing Qubit and Entanglement Error Rates}

\begin{table*}[ht]
\small
  \caption{Resource estimates for high ($10^{-3}$) error rate qubit architectures with varied node sizes. Bell state error rate is assumed to be $5\%$ and entanglement rate to be $10MHz$ throughout.}
  \centering
  \scalebox{0.7}{
  \begin{tabular}{lrlrlrlr}
    \toprule
    \multirow{2}{*}{\textbf{Application}} &
      \multicolumn{2}{c}{\textbf{45k qubit node}} &
      \multicolumn{2}{c}{\textbf{75k qubit node}} &
      \multicolumn{2}{c}{\textbf{100k qubit node}} \\
                          & Qubits  & Runtime    & Qubits  & Runtime    & Qubits  & Runtime\\
    \midrule
    Ising 10x10           & 0.38M & 1.1 min   & 0.41M & 0.94 min   & 0.56M & 0.47 min    \\ 
    Fermi-Hubbard 10x10   & 2.3M  & 12 hours   & 1.5M  & 11 hours  & 1.8M  & 6.8 hours   \\ 
    Heisenberg 10x10      & 1.4M  & 18 days  & 0.96M  & 18 days  & 0.85M  & 17 days   \\ 
    \midrule
    Shor's Factoring 2048 & 130M   & 9.9 days    & 74M   & 9.6 days & 66M   & 9.3 days   \\ 
    \midrule
    ZnS QPE               & 3.6M  & 3.0 months  & 3.1M  & 1.5 months  & 3.0M  & 1.4 months   \\ 
    Benzene QPE           & 7.8M  & 16 months  & 4.7M  & 7.5 months  & 4.3M   & 6.3 months \\ 
    Ruthenium QPE         & 22M   & 15 months  & 15M   & 7.3 months  & 12M   & 7.3 months \\ 
    Nitrogenase QPE       & ---   & --- & 26M   & 16 years & 22M   & 16 years \\ 
    \bottomrule
  \end{tabular}
  }
  \label{tab:estimatehigherrorrate}
\end{table*}

Comparing $10^{-3}$ and $10^{-4}$ physical error rate configurations in Figure \ref{fig:nodesize} and Figure \ref{fig:entanglementrate}, overheads of distributed computation are significantly lower with $10^{-4}$ qubits. 
It is well known that physical gate error rates are a massive contributor to the cost of quantum computation. In monolithic setting, higher error rates require higher distance surface code to perform computation fault tolerantly, increasing both qubit requirements and runtime. As we see from above simulations, the contribution of higher error rate is even worse in the distributed setting, as on top of impacting the resource requirements of the local parts of the algorithm, the distribution overhead also increases. This is because the EDFs on a higher error rate device need to use a higher distance of the underlying surface code in order to correct the errors. Due of this, entanglement distillation occupies a higher proportion of the node, leaving less space for compute qubits. Table \ref{tab:estimatehigherrorrate} contains resource requirements for a high error rate architecture, which are prohibitively large for all applications other than early quantum dynamics demonstrations. We therefore suggest that reducing the error rate of physical operations below $10^{-4}$is paramount to (distributed) quantum computing viability.

Entanglement error rate affects the resource requirements of the computation, but to a lesser degree. We compare fast qubits with 1\%  and   5\% Bell error rate in Figure \ref{fig:entanglementrate}. Lower error rate leads to significant improvements at the lower range of entanglement generation rate. But, once saturation is achieved, we get modest to no improvement. This can be explained by the fact that repetition code EDFs which are selected by our estimator for this regime, are extremely spacetime efficient at distilling high error rate states. Bringing 5\% infidelity Bell pair to 1\% infidelity only takes 4 physical qubits and 4 layers of physical operations using repetition codes; doing so, however, would require 4$\times$ more inputs, making the conversion undesirable when at high $\eta$. This shows that at high error rates between 0.1-5\% we can easily trade-off $\eta$ and Bell error rate.

Across application, geomean space time volume overhead over monolithic systems is 3.4$\times$ with 1\% Bell error rate and 3.1$\times$ with 0.1\% Bell error rate. Therefore, 1\% Bell error rate is a good goal for future hardware. DARPA's \cite{DARPA} target of 0.1\% Bell state error rate may not be required in practice. 
\subsection{System organization}

Across configurations, optimal space–time volumes were achieved when 25-30\% (max. 64\%) of qubits are dedicated to EDFs.
This fraction is strongly dependent on both node size and entanglement generation rate. At lower generation rates, fewer factories can operate at capacity, freeing more qubits for computation at the expense of runtime. Conversely, in larger nodes, the fixed number of networking qubits constitutes a smaller proportion of the total, allowing a greater allocation to computation. This proportion also indicates that linear connectivity is the ideal choice for practical distributed systems --- EDFs occupy a significant part of the node even with linear connectivity. 

The fraction of qubits required for magic state distillation is highly application-specific. For workloads with a relatively low runtime-to-qubit ratio, such as integer factoring, as little as 1.5\% of non-networking qubits are allocated to magic state distillation. In contrast, applications with long runtimes, such as ground-state estimation, require up to 40\% of non-networking qubits for magic state distillation.

\subsection{Early Prospects of Quantum Advantage}

Small applications typically have low networking overheads. Large applications inherently require more Bell pairs, necessitating higher Bell-pair fidelity and larger space-time volume to satisfy error thresholds. For early quantum advantage demonstrations with less than 0.5M qubits, distributed quantum architectures have geomean 3-4$\times$ space-time overhead over monolithic. However, these only require individual nodes with 15-25K qubits as opposed to a single node with up to 0.5M qubits in the monolithic case. For commercially relevant applications in quantum chemistry, geomean overheads increase to 5-8$\times$, but demand 45K node sizes.  

We note that even though our resource estimation framework and architecture are designed for superconducting qubit hardware, any device that is able to perform lattice surgery based operations can be modelled. Neutral Atom quantum computers have been shown to scale to much greater sizes than superconducting computers and although their slower operation times make them incapable of running complex algorithms such as Factoring and Quantum Chemistry in a reasonable timeframe, they are the most likely candidate for early quantum advantage demonstrations. 

In table ~\ref{tab:early-ising} we present resource estimates for running 10x10 Ising model on early distributed neutral atom quantum computers using lattice surgery. 
Our estimates demonstrate that if local gate fidelity of $10^{-4}$ can be achieved, then quantum advantage is possible with devices with as few as 5000 qubits. Monolithic neutral atom systems are already approaching these system sizes~\cite{Lukin3000, AC1000}. If they are paired with appropriate interconnects~\cite{Sunami2025}, this may be a promising direction for early quantum advantage. 

While neutral atoms supports native transversal gates which could further reduce operation times, our results demonstrate that even under the more constrained lattice surgery model, practical quantum advantage is achievable. These results provides a conservative upper bound on resource requirements for this platform.
 

\begin{table}[ht]
\small
  \centering
  \caption{Early-stage Ising model hardware configurations on slow ($100\mu$s) qubit architectures with $10^{-4}$ error rate}
  \begin{tabular}{@{} l  c  c  c  c @{}}
    \toprule
    \makecell{Node\\ Size} & \makecell{Number of \\nodes used} & \makecell{Entanglement \\rate} & \makecell{Total\\ qubits} & Runtime \\
    \midrule
    \makecell[l]{5000} &
      \makecell{26 \\ 27} &
      \makecell{4 kHz \\ 4 kHz} &
      \makecell{126k k \\ 131 k} &
      \makecell{46.8 hr \\ 25 hr} \\
    \addlinespace
    \makecell[l]{15000} &
      \makecell{6 \\ 7} &
      \makecell{4 kHz \\  10 kHz} &
      \makecell{83.5 k \\ 107 k} &
      \makecell{23.4 hr \\ 12.7 hr} \\
    \addlinespace
    \makecell[l]{25000} &
      \makecell{4 \\ 5} &
      \makecell{4 kHz \\  10 kHz} &
      \makecell{89.2 k \\ 121 k} &
      \makecell{13.1 hr \\ 7.8 hr} \\
    \addlinespace
    \makecell[l]{25000} &
      \makecell{4 \\ 5} &
      \makecell{4 kHz \\  10 kHz} &
      \makecell{89.2 k \\ 121 k} &
      \makecell{13.1 hr \\ 7.8 hr} \\
    \makecell[l]{45000} &
      \makecell{2 \\ 2} &
      \makecell{4 kHz \\  10 kHz} &
      \makecell{87.7 k \\ 84.5 k} &
      \makecell{11.7 hr \\ 7.8 hr} \\
    \addlinespace
    \makecell[l]{91542} &
      1 &
      --- &
      91.5 k &
      4.5 hr \\
    \bottomrule
  \end{tabular}
  \label{tab:early-ising}
\end{table}

\section{Related Work}


\textbf{Monolithic resource estimation:} Resource estimation has been studied for monolithic systems. Some examples are ~\cite{1611.07995, gidney2021factor}, which provide manual estimates. Automated frameworks have emerged for resource estimation, including Azure Quantum Resource Estimator~\cite{Beverland2022-kd}, Google Qualtran~\cite{harrigan2024qualtran} and BenchQ~\cite{BenchQ}. These works do not model distributed quantum computation, we leverage their ideas for modelling resource usage within a node in a distributed system. 

\textbf{Distributed quantum architecture:} Rigetti Resource Estimator \cite{2406.06015} attempts to model distributed quantum computation at a low level considering power consumption and heat dissipation of operations and components. It does not, however, include space needs for entanglement distillation modelling or associated errors (it only assumes that inter-module operations take $1 \mu s$ irrespective of fidelity). Our work is first to perform accurate modelling of entanglement distillation, revealing that 30-64\% of the qubits would need to be allocated for this component. When comparing the two tools on a 10x10 Ising model benchmark with similar hardware parameters ($10^{-3} $ physical operation error rate and $2$M qubit nodes), Rigetti RE estimated the need for $17.4$M qubits and $5.07$ days. Our tool estimated $1.06$M qubits and $20$ seconds. We expect that this is due to our diligent use of magic state factories and optimized entanglement distillation. Due to challenges with using the code for Rigetti RE we were unable to conduct a more rigorous comparison.

\cite{sinclair2025fault,Ramette2024} propose using Bell pairs to replace physical operations involved in lattice surgery remotely between logical qubits. 

This requires Bell pair fidelity to be on par with physical operation fidelity. They focus solely on the logical qubit interface and do not provide resource estimates for applications. 

\cite{10.1145/3620665.3640388} develops micro-architecture for distributed FTQC systems. It is focused on control architecture and applies only to superconducting qubits.

\textbf{Compilation:} 
Several works have studied applications mappings onto distributed architectures \cite{Wu2023, Feng2024, Cuomo2023, Eisert2000, AndresMartinez2024, AndrsMartnez2019}. These works operate on general quantum circuits without applying Clifford optimization~\cite{Litinski2019}, leading to very inefficient compiled programs with high gate counts. 

\section{Conclusion}
With monolithic quantum architectures facing severe scaling limits, distributed computing is emerging as the path to practical quantum computing. We develop one of the first resource estimators for distributed quantum computers. Analysing applications across hardware regimes, we provide important insights on node sizing, target error rates and entanglement generation rates and overall system organization. Our work shows that distributed quantum computing is practical both short-term as an avenue for early scientific demonstrations and long-term for commercial applications. In short-term, node sizes of 5000 qubits are sufficient to demonstrate quantum advantage. In long term, by scaling up current quantum computers to around 40K qubits while also focusing on building a performant quantum network, we can scale out to distributed systems that can target practical quantum advantage. We will open-source our resource estimator to enable further studies.

\bibliographystyle{ACM-Reference-Format}
\bibliography{refs}
1
\appendix
\label{rewriterule}
\section{Rewrite rule correctness proof}

In order to prove correctness of the rewrite rule in figure \ref{fig:nnodepauli} we will use ZX-calculus \cite{Cowtan2020}, a graphical language designed for quantum circuit rewrites. MQPRs are equivalent to what is known in ZX-calculus as Phase Gadgets, 

\begin{equation}\label{gadget}
\begin{tikzpicture}
	\begin{pgfonlayer}{nodelayer}
		\node [style=none] (1) at (1.5, -1) {};
		\node [style=none] (2) at (1.5, 0) {};
		\node [style=none] (3) at (1.5, 1) {};
		\node [style=none] (46) at (0, 1) {};
		\node [style=none] (47) at (0, 0) {};
		\node [style=none] (48) at (0, -1) {};
		\node [style=none] (50) at (0, -0.5) {$\vdots$};
		\node [style=none] (51) at (0.5, 1.25) {};
		\node [style=none] (52) at (1.5, 1.25) {};
		\node [style=none] (53) at (0.5, -1.25) {};
		\node [style=none] (54) at (1.5, -1.25) {};
		\node [style=none, rotate=90] (55) at (1, 0) {$P(\alpha)$};
		\node [style=none] (56) at (2, -1) {};
		\node [style=none] (57) at (2, 0) {};
		\node [style=none] (58) at (2, 1) {};
		\node [style=none] (59) at (0.5, -1) {};
		\node [style=none] (60) at (0.5, 0) {};
		\node [style=none] (61) at (0.5, 1) {};
		\node [style=none] (62) at (2, -0.5) {$\vdots$};
	\end{pgfonlayer}
	\begin{pgfonlayer}{edgelayer}
		\draw (51.center) to (53.center);
		\draw (53.center) to (54.center);
		\draw (54.center) to (52.center);
		\draw (52.center) to (51.center);
		\draw (46.center) to (61.center);
		\draw (47.center) to (60.center);
		\draw (48.center) to (59.center);
		\draw (1.center) to (56.center);
		\draw (2.center) to (57.center);
		\draw (3.center) to (58.center);
	\end{pgfonlayer}
\end{tikzpicture} = \begin{tikzpicture}
	\begin{pgfonlayer}{nodelayer}
		\node [style=none] (1) at (0, 1) {};
		\node [style=none] (3) at (0, 0.5) {};
		\node [style=none] (4) at (0, -0.5) {};
		\node [style=X dot] (8) at (3, -1) {};
		\node [style=Z phase dot] (9) at (4, -1) {$a$};
		\node [style=none] (14) at (4.5, 0.5) {};
		\node [style=none] (16) at (4.5, 1) {};
		\node [style=none] (17) at (4.5, -0.5) {};
		\node [style=Z box] (18) at (1.5, 1) {$P_1$};
		\node [style=Z box] (20) at (1.5, 0.5) {$P_2$};
        \node [style=none] (62) at (1.5, 0) {$\vdots$};
		\node [style=Z box] (21) at (1.5, -0.5) {$P_n$};
	\end{pgfonlayer}
	\begin{pgfonlayer}{edgelayer}
		\draw (1.center) to (18);
		\draw (3.center) to (20);
		\draw (4.center) to (21);
		\draw (8) to (9);
		\draw (8) to (20);
		\draw (8) to (21);
		\draw (8) to (18);
		\draw (14.center) to (20);
		\draw (16.center) to (18);
		\draw (17.center) to (21);
	\end{pgfonlayer}
\end{tikzpicture}
\end{equation}

For example, 

\begin{equation}\label{IXYZGadget}
e^{-i\alpha I X Y Z} =  {\begin{tikzpicture}
	\begin{pgfonlayer}{nodelayer}
		\node [style=none] (1) at (0, 1) {};
		\node [style=none] (2) at (0, 0.5) {};
		\node [style=none] (3) at (0, -0.5) {};
		\node [style=none] (4) at (0, -1) {};
		\node [style=hadamard] (5) at (1, 0.5) {};
		\node [style=Z dot] (6) at (2, 0.5) {};
		\node [style=hadamard] (7) at (4.5, 0.5) {};
		\node [style=X dot] (8) at (2.5, 0) {};
		\node [style=Z phase dot] (9) at (3.5, 0) {$2\alpha$};
		\node [style=Z dot] (10) at (2, -0.5) {};
		\node [style=Z dot] (11) at (2, -1) {};
		\node [style=X phase dot] (12) at (1, -0.5) {$+$};
		\node [style=X phase dot] (13) at (3, -0.5) {$-$};
		\node [style=none] (14) at (5.5, -0.5) {};
		\node [style=none] (15) at (5.5, 0.5) {};
		\node [style=none] (16) at (5.5, 1) {};
		\node [style=none] (17) at (5.5, -1) {};
	\end{pgfonlayer}
	\begin{pgfonlayer}{edgelayer}
		\draw (1.center) to (16.center);
		\draw (2.center) to (5);
		\draw (3.center) to (12);
		\draw (4.center) to (11);
		\draw (5) to (6);
		\draw (6) to (8);
		\draw (6) to (7);
		\draw (7) to (15.center);
		\draw (8) to (10);
		\draw (8) to (11);
		\draw (8) to (9);
		\draw (10) to (12);
		\draw (10) to (13);
		\draw (11) to (17.center);
		\draw (13) to (14.center);
	\end{pgfonlayer}
\end{tikzpicture}}
\end{equation}

Lattice surgery implements the Pauli boxes directly so from here on, we will use the representation like in equation \ref{IXYZGadget} but ignoring the "basis change" gates surrounding the gadget.

In practice, gadgets are implemented using state injection. Measurements during the state injection protocol introduce random errors to the rotation, but these errors can be corrected using Pauli gadgets which don't need to be executed on quantum hardware and can simply be commuted to the end of the circuit. Here, an error laBelled as $a\pi$ could be resolved with a $\pi$ gadget of the same Pauli string and $b\pi$ can be resolved with the $\pi/2$ rotation \cite{LitinskiMajoranaFermion, Litinski2019}

\begin{equation}\label{stateinjectiongadget}
\begin{tikzpicture}
    \begin{pgfonlayer}{nodelayer}
        \node [style=none] (0) at (1.00, 0.50) {};
        \node [style=none] (1) at (1.00, -0.50) {};
        \node [style=none] (2) at (1.00, -2.00) {};
        \node [style=none] (3) at (1.00, -3.00) {};
        \node [style=none] (20) at (6.00, 0.50) {};
        \node [style=none] (21) at (6.00, -0.50) {};
        \node [style=none] (22) at (6.00, -2.00) {};
        \node [style=none] (23) at (6.00, -3.00) {};
        \node [style=Z dot] (24) at (2.50, 0.50) {};
        \node [style=Z dot] (25) at (2.50, -0.50) {};
		\node [style=none] (50) at (2.50, -1) {$\vdots$};
        \node [style=Z dot] (26) at (2.50, -2.00) {};
        \node [style=Z dot] (27) at (2.50, -3.00) {};
        \node [style=X dot] (28) at (3.50, -4.00) {};
        \node [style=Z phase dot] (29) at (1.50, -5.00) {$ \pi/4$};
        \node [style=X dot] (30) at (1.50, -4.00) {};
        \node [style=X phase dot] (31) at (5.50, -4.00) {$b\pi$};
        \node [style=Z phase dot] (32) at (5.50, -5.00) {$a\pi$};
        \node [style=Z dot] (33) at (3.50, -5.00) {};
    \end{pgfonlayer}
    \begin{pgfonlayer}{edgelayer}
        \draw (0) to (24);
        \draw (1) to (25);
        \draw (2) to (26);
        \draw (3) to (27);
        \draw (20) to (24);
        \draw (21) to (25);
        \draw (22) to (26);
        \draw (23) to (27);
        \draw (24) to (28);
        \draw (25) to (28);
        \draw (26) to (28);
        \draw (27) to (28);
        \draw (28) to (30);
        \draw (28) to (31);
        \draw (28) to (33);
        \draw (29) to (33);
        \draw (32) to (33);
    \end{pgfonlayer}
\end{tikzpicture} = \begin{tikzpicture}
    \begin{pgfonlayer}{nodelayer}
        \node [style=none] (0) at (1.00, 0.50) {};
        \node [style=none] (1) at (1.00, -0.50) {};
        \node [style=none] (2) at (1.00, -2.00) {};
        \node [style=none] (3) at (1.00, -3.00) {};
        \node [style=none] (20) at (6.00, 0.50) {};
        \node [style=none] (21) at (6.00, -0.50) {};
        \node [style=none] (22) at (6.00, -2.00) {};
        \node [style=none] (23) at (6.00, -3.00) {};
        \node [style=Z dot] (24) at (2.50, 0.50) {};
        \node [style=Z dot] (25) at (2.50, -0.50) {};
		\node [style=none] (50) at (2.50, -1) {$\vdots$};
        \node [style=Z dot] (26) at (2.50, -2.00) {};
        \node [style=Z dot] (27) at (2.50, -3.00) {};
        \node [style=X phase dot] (28) at (3.50, -4.00) {$b\pi$};
        \node [style=Z phase dot] (29) at (5.00, -4.00) {$\pi/4 + a\pi$};
    \end{pgfonlayer}
    \begin{pgfonlayer}{edgelayer}
        \draw (0) to (24);
        \draw (1) to (25);
        \draw (2) to (26);
        \draw (3) to (27);
        \draw (20) to (24);
        \draw (21) to (25);
        \draw (22) to (26);
        \draw (23) to (27);
        \draw (24) to (28);
        \draw (25) to (28);
        \draw (26) to (28);
        \draw (27) to (28);
        \draw (28) to (29);
    \end{pgfonlayer}
\end{tikzpicture} = \begin{tikzpicture}
    \begin{pgfonlayer}{nodelayer}
        \node [style=none] (0) at (1.00, 0.50) {};
        \node [style=none] (1) at (1.00, -0.50) {};
        \node [style=none] (2) at (1.00, -2.00) {};
        \node [style=none] (3) at (1.00, -3.00) {};
        \node [style=none] (20) at (6.00, 0.50) {};
        \node [style=none] (21) at (6.00, -0.50) {};
        \node [style=none] (22) at (6.00, -2.00) {};
        \node [style=none] (23) at (6.00, -3.00) {};
        \node [style=Z dot] (24) at (2.50, 0.50) {};
        \node [style=Z dot] (25) at (2.50, -0.50) {};
		\node [style=none] (50) at (2.50, -1) {$\vdots$};
        \node [style=Z dot] (26) at (2.50, -2.00) {};
        \node [style=Z dot] (27) at (2.50, -3.00) {};
        \node [style=X dot] (28) at (3.50, -4.00) {};
        \node [style=Z phase dot] (29) at (6.00, -4.00) {$(-1)^{b} (\pi/4 + a\pi)$};
    \end{pgfonlayer}
    \begin{pgfonlayer}{edgelayer}
        \draw (0) to (24);
        \draw (1) to (25);
        \draw (2) to (26);
        \draw (3) to (27);
        \draw (20) to (24);
        \draw (21) to (25);
        \draw (22) to (26);
        \draw (23) to (27);
        \draw (24) to (28);
        \draw (25) to (28);
        \draw (26) to (28);
        \draw (27) to (28);
        \draw (28) to (29);
    \end{pgfonlayer}
\end{tikzpicture}
\end{equation}

Now, in the distributed setting we utilise Bell states in order to entangle the ancillae of different nodes through teleportation. We present an example with 3 nodes, but the connection between the top two nodes can be replicated any number of times in order to perform a gadget on an arbitrary number of nodes. Note that in figure \ref{fig:nnodepauli} we represented entanglement of the Bell states and ancilla with an extension of the Multi-Qubit Pauli Measurement to include Bell state measured in the X basis. This is not accurate, since this Pauli measurement would include a CX gate implemented with a twist. In reality a "merge" operation is performed instead between the ancilla nad Bell state, which is still native to the surface code. This choice of representation in figure \ref{fig:nnodepauli} is chosen for conciseness as it communicates the idea of the operation being performed.

\begin{equation}\label{stateinjectiongadget}
\begin{split}
\begin{tikzpicture}
    \begin{pgfonlayer}{nodelayer}
        \node [style=none] (0) at (1.00, 1.50) {};
        \node [style=none] (1) at (1.00, -0.00) {};
        \node [style=none] (2) at (1.00, -2.00) {};
        \node [style=none] (3) at (1.00, -3.00) {};
        \node [style=none] (20) at (6.00, 1.50) {};
        \node [style=none] (21) at (6.00, -0.00) {};
        \node [style=none] (22) at (6.00, -2.00) {};
        \node [style=none] (23) at (6.00, -3.00) {};
        \node [style=Z dot] (24) at (2.50, 1.50) {};
        \node [style=Z dot] (25) at (2.50, -0.00) {};
        \node [style=Z dot] (26) at (2.50, -2.00) {};
        \node [style=Z dot] (27) at (2.50, -3.00) {};
        \node [style=X dot] (28) at (3.50, -4.00) {};
        \node [style=Z phase dot] (29) at (1.50, -4.50) {$\pi/4$};
        \node [style=X dot] (30) at (1.50, -4.00) {};
        \node [style=X phase dot] (31) at (5.50, -4.00) {$a_7\pi$};
        \node [style=Z phase dot] (32) at (5.50, -4.50) {$b\pi$};
        \node [style=Z dot] (33) at (3.50, -4.50) {};
        \node [style=Z dot] (34) at (2.50, 5.00) {};
        \node [style=none] (35) at (6.00, 3.50) {};
        \node [style=Z dot] (36) at (2.50, 3.50) {};
        \node [style=none] (37) at (6.00, 5.00) {};
        \node [style=none] (38) at (1.00, 5.00) {};
        \node [style=none] (39) at (1.00, 3.50) {};
        \node [style=X dot] (40) at (4.00, 3.00) {};
        \node [style=X dot] (41) at (4.00, 2.00) {};
        \node [style=X dot] (42) at (4.00, 4.25) {};
        \node [style=X dot] (43) at (4.00, 0.75) {};
        \node [style=X dot] (44) at (4.00, -0.50) {};
        \node [style=X dot] (45) at (4.00, -1.50) {};
        \node [style=X phase dot] (46) at (5.00, 4.25) {$a_1\pi$};
        \node [style=X phase dot] (47) at (5.00, 3.00) {$a_2\pi$};
        \node [style=X phase dot] (48) at (5.00, 2.00) {$a_3\pi$};
        \node [style=X phase dot] (49) at (5.00, 0.75) {$a_4\pi$};
        \node [style=X phase dot] (50) at (5.00, -0.50) {$a_5\pi$};
        \node [style=X phase dot] (51) at (5.00, -1.50) {$a_6\pi$};
        \node [style=Z dot] (54) at (2.00, -0.50) {};
        \node [style=Z dot] (55) at (2.00, -1.50) {};
        \node [style=Z dot] (52) at (2.00, 3.00) {};
        \node [style=Z dot] (53) at (2.00, 2.00) {};
    \end{pgfonlayer}
    \begin{pgfonlayer}{edgelayer}
        \draw (0) to (24);
        \draw (1) to (25);
        \draw (2) to (26);
        \draw (3) to (27);
        \draw (20) to (24);
        \draw (21) to (25);
        \draw (22) to (26);
        \draw (23) to (27);
        \draw (24) to (43);
        \draw (25) to (43);
        \draw (26) to (28);
        \draw (27) to (28);
        \draw (28) to (30);
        \draw (28) to (31);
        \draw (28) to (33);
        \draw (28) to (45);
        \draw (29) to (33);
        \draw (32) to (33);
        \draw (34) to (37);
        \draw (34) to (38);
        \draw (34) to (42);
        \draw (35) to (36);
        \draw (36) to (39);
        \draw (36) to (42);
        \draw (40) to (42);
        \draw (40) to (47);
        \draw (40) to (52);
        \draw (41) to (43);
        \draw (41) to (48);
        \draw (41) to (53);
        \draw (42) to (46);
        \draw (43) to (44);
        \draw (43) to (49);
        \draw (44) to (50);
        \draw (44) to (54);
        \draw (45) to (51);
        \draw (45) to (55);
        \draw[very thick] (54) to (55); 
        \draw[very thick] (52) to (53);
    \end{pgfonlayer}
\end{tikzpicture} = \begin{tikzpicture}
    \begin{pgfonlayer}{nodelayer}
        \node [style=none] (0) at (1.00, 1.50) {};
        \node [style=none] (1) at (1.00, -0.00) {};
        \node [style=none] (2) at (1.00, -2.00) {};
        \node [style=none] (3) at (1.00, -3.00) {};
        \node [style=none] (20) at (6.00, 1.50) {};
        \node [style=none] (21) at (6.00, -0.00) {};
        \node [style=none] (22) at (6.00, -2.00) {};
        \node [style=none] (23) at (6.00, -3.00) {};
        \node [style=Z dot] (24) at (2.50, 1.50) {};
        \node [style=Z dot] (25) at (2.50, -0.00) {};
        \node [style=Z dot] (26) at (2.50, -2.00) {};
        \node [style=Z dot] (27) at (2.50, -3.00) {};
        \node [style=X dot] (28) at (3.50, -4.00) {};
        \node [style=Z phase dot] (29) at (1.50, -4.50) {$\pi/4$};
        \node [style=X dot] (30) at (1.50, -4.00) {};
        \node [style=X phase dot] (31) at (5.50, -4.00) {$a_7\pi$};
        \node [style=Z phase dot] (32) at (5.50, -4.50) {$b\pi$};
        \node [style=Z dot] (33) at (3.50, -4.50) {};
        \node [style=Z dot] (34) at (2.50, 5.00) {};
        \node [style=none] (35) at (6.00, 3.50) {};
        \node [style=Z dot] (36) at (2.50, 3.50) {};
        \node [style=none] (37) at (6.00, 5.00) {};
        \node [style=none] (38) at (1.00, 5.00) {};
        \node [style=none] (39) at (1.00, 3.50) {};
        \node [style=X dot] (40) at (4.00, 3.00) {};
        \node [style=X dot] (41) at (4.00, 2.00) {};
        \node [style=X dot] (42) at (4.00, 4.25) {};
        \node [style=X dot] (43) at (4.00, 0.75) {};
        \node [style=X dot] (44) at (4.00, -0.50) {};
        \node [style=X dot] (45) at (4.00, -1.50) {};
        \node [style=X phase dot] (46) at (5.00, 4.25) {$a_1\pi$};
        \node [style=X phase dot] (47) at (5.00, 3.00) {$a_2\pi$};
        \node [style=X phase dot] (48) at (5.00, 2.00) {$a_3\pi$};
        \node [style=X phase dot] (49) at (5.00, 0.75) {$a_4\pi$};
        \node [style=X phase dot] (50) at (5.00, -0.50) {$a_5\pi$};
        \node [style=X phase dot] (51) at (5.00, -1.50) {$a_6\pi$};
    \end{pgfonlayer}
    \begin{pgfonlayer}{edgelayer}
        \draw (0) to (24);
        \draw (1) to (25);
        \draw (2) to (26);
        \draw (3) to (27);
        \draw (20) to (24);
        \draw (21) to (25);
        \draw (22) to (26);
        \draw (23) to (27);
        \draw (24) to (43);
        \draw (25) to (43);
        \draw (26) to (28);
        \draw (27) to (28);
        \draw (28) to (30);
        \draw (28) to (31);
        \draw (28) to (33);
        \draw (28) to (45);
        \draw (29) to (33);
        \draw (32) to (33);
        \draw (34) to (37);
        \draw (34) to (38);
        \draw (34) to (42);
        \draw (35) to (36);
        \draw (36) to (39);
        \draw (36) to (42);
        \draw (40) to (42);
        \draw (40) to (47);
        \draw (40) to (41);
        \draw (41) to (43);
        \draw (41) to (48);
        \draw (42) to (46);
        \draw (43) to (44);
        \draw (43) to (49);
        \draw (44) to (50);
        \draw (44) to (45);
        \draw (45) to (51);
    \end{pgfonlayer}
\end{tikzpicture} = \begin{tikzpicture}
    \begin{pgfonlayer}{nodelayer}
        \node [style=none] (0) at (1.00, 1.50) {};
        \node [style=none] (1) at (1.00, -0.00) {};
        \node [style=none] (2) at (1.00, -2.00) {};
        \node [style=none] (3) at (1.00, -3.00) {};
        \node [style=none] (20) at (6.00, 1.50) {};
        \node [style=none] (21) at (6.00, -0.00) {};
        \node [style=none] (22) at (6.00, -2.00) {};
        \node [style=none] (23) at (6.00, -3.00) {};
        \node [style=Z dot] (24) at (2.50, 1.50) {};
        \node [style=Z dot] (25) at (2.50, -0.00) {};
        \node [style=Z dot] (26) at (2.50, -2.00) {};
        \node [style=Z dot] (27) at (2.50, -3.00) {};
        \node [style=X phase dot] (28) at (3.50, -4.00) {$a_7\pi$};
        \node [style=Z phase dot] (33) at (5, -4.00) {$\pi/4 + b\pi$};
        \node [style=Z dot] (34) at (2.50, 5.00) {};
        \node [style=none] (35) at (6.00, 3.50) {};
        \node [style=Z dot] (36) at (2.50, 3.50) {};
        \node [style=none] (37) at (6.00, 5.00) {};
        \node [style=none] (38) at (1.00, 5.00) {};
        \node [style=none] (39) at (1.00, 3.50) {};
        \node [style=X dot] (40) at (4.00, 3.00) {};
        \node [style=X dot] (41) at (4.00, 2.00) {};
        \node [style=X dot] (42) at (4.00, 4.25) {};
        \node [style=X dot] (43) at (4.00, 0.75) {};
        \node [style=X dot] (44) at (4.00, -0.50) {};
        \node [style=X dot] (45) at (4.00, -1.50) {};
        \node [style=X phase dot] (46) at (5.00, 4.25) {$a_1\pi$};
        \node [style=X phase dot] (47) at (5.00, 3.00) {$a_2\pi$};
        \node [style=X phase dot] (48) at (5.00, 2.00) {$a_3\pi$};
        \node [style=X phase dot] (49) at (5.00, 0.75) {$a_4\pi$};
        \node [style=X phase dot] (50) at (5.00, -0.50) {$a_5\pi$};
        \node [style=X phase dot] (51) at (5.00, -1.50) {$a_6\pi$};
    \end{pgfonlayer}
    \begin{pgfonlayer}{edgelayer}
        \draw (0) to (24);
        \draw (1) to (25);
        \draw (2) to (26);
        \draw (3) to (27);
        \draw (20) to (24);
        \draw (21) to (25);
        \draw (22) to (26);
        \draw (23) to (27);
        \draw (24) to (43);
        \draw (25) to (43);
        \draw (26) to (28);
        \draw (27) to (28);
        \draw (28) to (33);
        \draw (28) to (45);
        \draw (34) to (37);
        \draw (34) to (38);
        \draw (34) to (42);
        \draw (35) to (36);
        \draw (36) to (39);
        \draw (36) to (42);
        \draw (40) to (42);
        \draw (40) to (47);
        \draw (40) to (41);
        \draw (41) to (43);
        \draw (41) to (48);
        \draw (42) to (46);
        \draw (43) to (44);
        \draw (43) to (49);
        \draw (44) to (50);
        \draw (44) to (45);
        \draw (45) to (51);
    \end{pgfonlayer}
\end{tikzpicture} =
\\
 \begin{tikzpicture}
    \begin{pgfonlayer}{nodelayer}
        \node [style=none] (0) at (1.00, 1.50) {};
        \node [style=none] (1) at (1.00, -0.00) {};
        \node [style=none] (2) at (1.00, -2.00) {};
        \node [style=none] (3) at (1.00, -3.00) {};
        \node [style=none] (20) at (6.00, 1.50) {};
        \node [style=none] (21) at (6.00, -0.00) {};
        \node [style=none] (22) at (6.00, -2.00) {};
        \node [style=none] (23) at (6.00, -3.00) {};
        \node [style=Z dot] (24) at (2.50, 1.50) {};
        \node [style=Z dot] (25) at (2.50, -0.00) {};
        \node [style=Z dot] (26) at (2.50, -2.00) {};
        \node [style=Z dot] (27) at (2.50, -3.00) {};
        \node [style=X phase dot] (28) at (4.50, 0.75) {$\pi\bigoplus_{0<i<8}a_i $};
        \node [style=Z phase dot] (33) at (7, 0.75) {$\pi/4 + b\pi$};
        \node [style=Z dot] (34) at (2.50, 5.00) {};
        \node [style=none] (35) at (6.00, 3.50) {};
        \node [style=Z dot] (36) at (2.50, 3.50) {};
        \node [style=none] (37) at (6.00, 5.00) {};
        \node [style=none] (38) at (1.00, 5.00) {};
        \node [style=none] (39) at (1.00, 3.50) {};
    \end{pgfonlayer}
    \begin{pgfonlayer}{edgelayer}
        \draw (0) to (24);
        \draw (1) to (25);
        \draw (2) to (26);
        \draw (3) to (27);
        \draw (20) to (24);
        \draw (21) to (25);
        \draw (22) to (26);
        \draw (23) to (27);
        \draw (24) to (28);
        \draw (25) to (28);
        \draw (26) to (28);
        \draw (27) to (28);
        \draw (28) to (33);
        \draw (28) to (36);
        \draw (28) to (34);
        \draw (34) to (37);
        \draw (34) to (38);
        \draw (35) to (36);
        \draw (36) to (39);
    \end{pgfonlayer}
\end{tikzpicture} = \begin{tikzpicture}
    \begin{pgfonlayer}{nodelayer}
        \node [style=none] (0) at (1.00, 1.50) {};
        \node [style=none] (1) at (1.00, -0.00) {};
        \node [style=none] (2) at (1.00, -2.00) {};
        \node [style=none] (3) at (1.00, -3.00) {};
        \node [style=none] (20) at (6.00, 1.50) {};
        \node [style=none] (21) at (6.00, -0.00) {};
        \node [style=none] (22) at (6.00, -2.00) {};
        \node [style=none] (23) at (6.00, -3.00) {};
        \node [style=Z dot] (24) at (2.50, 1.50) {};
        \node [style=Z dot] (25) at (2.50, -0.00) {};
        \node [style=Z dot] (26) at (2.50, -2.00) {};
        \node [style=Z dot] (27) at (2.50, -3.00) {};
        \node [style=X dot] (28) at (4.50, 0.75) {};
        \node [style=Z phase dot] (33) at (7, 0.75) {$(-1)^{(\bigoplus a_i )}(\pi/4 + b\pi)$};
        \node [style=Z dot] (34) at (2.50, 5.00) {};
        \node [style=none] (35) at (6.00, 3.50) {};
        \node [style=Z dot] (36) at (2.50, 3.50) {};
        \node [style=none] (37) at (6.00, 5.00) {};
        \node [style=none] (38) at (1.00, 5.00) {};
        \node [style=none] (39) at (1.00, 3.50) {};
    \end{pgfonlayer}
    \begin{pgfonlayer}{edgelayer}
        \draw (0) to (24);
        \draw (1) to (25);
        \draw (2) to (26);
        \draw (3) to (27);
        \draw (20) to (24);
        \draw (21) to (25);
        \draw (22) to (26);
        \draw (23) to (27);
        \draw (24) to (28);
        \draw (25) to (28);
        \draw (26) to (28);
        \draw (27) to (28);
        \draw (28) to (33);
        \draw (28) to (36);
        \draw (28) to (34);
        \draw (34) to (37);
        \draw (34) to (38);
        \draw (35) to (36);
        \draw (36) to (39);
    \end{pgfonlayer}
\end{tikzpicture}
\end{split}
\end{equation}

The errors can then be corrected by $\pi$  and $\pi/4$  gadgets which are Clifford and so can be algorithmically commuted to the end of the circuit and absorbed into measurement, without impacting the use of quantum resources

\begin{equation}\label{stateinjectiongadget}
\begin{tikzpicture}
    \begin{pgfonlayer}{nodelayer}
        \node [style=none] (0) at (1.00, 1.50) {};
        \node [style=none] (1) at (1.00, -0.00) {};
        \node [style=none] (2) at (1.00, -2.00) {};
        \node [style=none] (3) at (1.00, -3.00) {};
        \node [style=none] (20) at (12.00, 1.50) {};
        \node [style=none] (21) at (12.00, -0.00) {};
        \node [style=none] (22) at (12.00, -2.00) {};
        \node [style=none] (23) at (12.00, -3.00) {};
        \node [style=Z dot] (24) at (2.50, 1.50) {};
        \node [style=Z dot] (25) at (2.50, -0.00) {};
        \node [style=Z dot] (26) at (2.50, -2.00) {};
        \node [style=Z dot] (27) at (2.50, -3.00) {};
        \node [style=X dot] (28) at (3, 0.75) {};
        \node [style=Z phase dot] (33) at (5.5, 0.75) {$(-1)^{(\bigoplus a_i )}(\pi/4 + b\pi)$};
        \node [style=Z dot] (34) at (2.50, 5.00) {};
        \node [style=none] (35) at (12.00, 3.50) {};
        \node [style=Z dot] (36) at (2.50, 3.50) {};
        \node [style=none] (37) at (12.00, 5.00) {};
        \node [style=none] (38) at (1.00, 5.00) {};
        \node [style=none] (39) at (1.00, 3.50) {};
        \node [style=X dot] (40) at (8, 0.75) {};
        \node [style=Z dot] (41) at (5.50, 3.50) {};
        \node [style=Z dot] (42) at (5.2, 1.50) {};
        \node [style=Z phase dot] (43) at (8.7, 0.75) {$b\pi$};
        \node [style=Z dot] (44) at (5.50, -3.00) {};
        \node [style=Z dot] (45) at (5.50, -0.00) {};
        \node [style=Z dot] (46) at (5.50, -2.00) {};
        \node [style=Z dot] (47) at (5.50, 5.00) {};
        \node [style=X dot] (48) at (10, 0.75) {};
        \node [style=Z dot] (49) at (8.50, 3.50) {};
        \node [style=Z dot] (50) at (8.50, 1.50) {};
        \node [style=Z phase dot] (51) at (11.5, 0.75) {$(\bigoplus a_i )\pi/2 $};
        \node [style=Z dot] (52) at (8.50, -3.00) {};
        \node [style=Z dot] (53) at (8.50, -0.00) {};
        \node [style=Z dot] (54) at (8.50, -2.00) {};
        \node [style=Z dot] (55) at (8.50, 5.00) {};
    \end{pgfonlayer}
    \begin{pgfonlayer}{edgelayer}
        \draw (0) to (24);
        \draw (1) to (25);
        \draw (2) to (26);
        \draw (3) to (27);
        \draw (20) to (24);
        \draw (21) to (25);
        \draw (22) to (26);
        \draw (23) to (27);
        \draw (24) to (28);
        \draw (25) to (28);
        \draw (26) to (28);
        \draw (27) to (28);
        \draw (28) to (33);
        \draw (28) to (36);
        \draw (28) to (34);
        \draw (34) to (37);
        \draw (34) to (38);
        \draw (35) to (36);
        \draw (36) to (39);
        \draw (40) to (43);
        \draw (40) to (41);
        \draw (40) to (47);
        \draw (40) to (42);
        \draw (40) to (44);
        \draw (40) to (45);
        \draw (40) to (46);
        \draw (48) to (51);
        \draw (48) to (49);
        \draw (48) to (55);
        \draw (48) to (50);
        \draw (48) to (52);
        \draw (48) to (53);
        \draw (48) to (54);
    \end{pgfonlayer}
\end{tikzpicture} = \begin{tikzpicture}
    \begin{pgfonlayer}{nodelayer}
        \node [style=none] (0) at (1.00, 1.50) {};
        \node [style=none] (1) at (1.00, -0.00) {};
        \node [style=none] (2) at (1.00, -2.00) {};
        \node [style=none] (3) at (1.00, -3.00) {};
        \node [style=none] (20) at (6.00, 1.50) {};
        \node [style=none] (21) at (6.00, -0.00) {};
        \node [style=none] (22) at (6.00, -2.00) {};
        \node [style=none] (23) at (6.00, -3.00) {};
        \node [style=Z dot] (24) at (2.50, 1.50) {};
        \node [style=Z dot] (25) at (2.50, -0.00) {};
        \node [style=Z dot] (26) at (2.50, -2.00) {};
        \node [style=Z dot] (27) at (2.50, -3.00) {};
        \node [style=X dot] (28) at (4.50, 0.75) {};
        \node [style=Z phase dot] (33) at (7, 0.75) {$\pi/4$};
        \node [style=Z dot] (34) at (2.50, 5.00) {};
        \node [style=none] (35) at (6.00, 3.50) {};
        \node [style=Z dot] (36) at (2.50, 3.50) {};
        \node [style=none] (37) at (6.00, 5.00) {};
        \node [style=none] (38) at (1.00, 5.00) {};
        \node [style=none] (39) at (1.00, 3.50) {};
    \end{pgfonlayer}
    \begin{pgfonlayer}{edgelayer}
        \draw (0) to (24);
        \draw (1) to (25);
        \draw (2) to (26);
        \draw (3) to (27);
        \draw (20) to (24);
        \draw (21) to (25);
        \draw (22) to (26);
        \draw (23) to (27);
        \draw (24) to (28);
        \draw (25) to (28);
        \draw (26) to (28);
        \draw (27) to (28);
        \draw (28) to (33);
        \draw (28) to (36);
        \draw (28) to (34);
        \draw (34) to (37);
        \draw (34) to (38);
        \draw (35) to (36);
        \draw (36) to (39);
    \end{pgfonlayer}
\end{tikzpicture}
\end{equation}

\section{Applications}
\noindent\textbf{Quantum Dynamics:} 
We benchmark quantum many-body dynamics using two complementary simulation strategies: a Trotterization-based approach \cite{lloyd1996universal}, which is the most commonly used algorithm for Hamiltonian simulation, for the transverse-field Ising model, and the time-optimal post-Trotter algorithm \cite{low2017optimal} for the Fermi-Hubbard and Heisenberg models. These models are canonical testbeds for quantum simulation, capturing essential features such as locality, noncommuting interactions, and tunable correlation strength. As benchmarks, they stress both circuit depth and gate scheduling while remaining physically well motivated. The post-Trotter method further incorporates advanced subroutines such as block encoding and quantum signal processing, making it representative of state-of-the-art Hamiltonian simulation algorithms. Together, these benchmarks probe the performance of fault-tolerant quantum processors on tasks that are classically intractable and central to condensed-matter physics.

\noindent\textbf{Molecular QPE:} 
We benchmark quantum chemistry workloads by estimating molecular ground-state energies using quantum phase estimation (QPE) \cite{babbush2018encoding}. QPE is a foundational quantum subroutine that appears in many advanced algorithms, and its inclusion makes this benchmark particularly informative. The workload combines several critical components---state preparation, controlled Hamiltonian simulation, and long coherent evolutions---thereby stressing both algorithmic depth and logical error rates. Medium-scale molecules such as benzene or catalytic clusters provide chemically relevant instances while remaining within plausible fault-tolerant resource envelopes, making this benchmark a realistic proxy for future quantum advantage in chemistry.

\noindent\textbf{Shor's Algorithm for Integer Factoring:} 
Shor's algorithm serves as a standard benchmark for fault-tolerant quantum computation, as it integrates multiple core subroutines including modular arithmetic, quantum Fourier transform, and phase estimation. Its polynomial-time scaling for integer factoring contrasts sharply with the best-known classical algorithms, making it a clear marker of quantum advantage. As a benchmark, Shor's algorithm exercises long-range entanglement, deep logical circuits, and error-corrected arithmetic, providing a stringent test of both architectural design and error-correction performance.

\end{document}